\documentclass[pdflatex,sn-nature]{sn-jnl}

\usepackage{graphicx}%
\usepackage{multirow}%
\usepackage{amsmath,amssymb,amsfonts}%
\usepackage{amsthm}%
\usepackage{mathrsfs}%
\usepackage[title]{appendix}%
\usepackage{xcolor}%
\usepackage{textcomp}%
\usepackage{manyfoot}%
\usepackage{booktabs}%
\usepackage{algorithm}%
\usepackage{algorithmicx}%
\usepackage{algpseudocode}%
\usepackage{listings}%

\theoremstyle{thmstyleone}%
\theoremstyle{thmstyletwo}%

\theoremstyle{thmstylethree}%

\usepackage{enumitem}
\usepackage{soul}
\usepackage{colortbl}
\usepackage{amsmath}
\usepackage{titlesec}

\newcommand{\tn}{\textnormal}

\begin{document}

\title[Magnetic Field-Mediated Superconducting Logic]{Magnetic Field-Mediated Superconducting Logic}

\author*[1,2]{\fnm{Alexander J.} \sur{Edwards}}\email{ajedwar@lps.umd.edu}
\equalcont{Alexander J. Edwards and Son T. Le contributed equally.}

\author*[1]{\fnm{Son T.} \sur{Le}}\email{stle@lps.umd.edu}
\equalcont{Alexander J. Edwards and Son T. Le contributed equally.}

\author[1]{\fnm{Nicholas W. G.} \sur{Smith}}

\author[2]{\fnm{Ebenezer C.} \sur{Usih}}

\author[1]{\fnm{Austin} \sur{Thomas}}

\author[1]{\fnm{Christopher J. K.} \sur{Richardson}}

\author[1]{\fnm{Nicholas A.} \sur{Blumenschein}}

\author[1]{\fnm{Aubrey T.} \sur{Hanbicki}}

\author*[1]{\fnm{Adam L.} \sur{Friedman}}\email{afriedman@lps.umd.edu}

\author*[2]{\fnm{Joseph S.} \sur{Friedman}}\email{Joseph.Friedman@utdallas.edu}

\affil[1]{\orgname{Laboratory for Physical Sciences}, \orgaddress{\city{College Park}, \state{MD}, \country{USA}}}

\affil[2]{\orgdiv{Department of Electrical and Computer Engineering}, \orgname{The University of Texas at Dallas}, \orgaddress{\city{Richardson}, \state{TX}, \country{USA}}}





\abstract{While superconductors are highly attractive for energy-efficient computing, fundamental limitations in their logic circuit integration have hindered scaling and led to increased energy consumption.  We therefore propose and experimentally demonstrate a novel superconducting switching device utilizing the proximity magnetization from a spin-orbit torque-switched magnet to control the resistivity of a superconductor.  We further propose a complete logic family comprised solely of these devices.  This novel implementation has the potential to drastically outperform existing superconducting logic families in terms of energy efficiency and scalability.}




\maketitle

\section{Introduction}

Superconducting logic is potentially orders of magnitude more efficient than state-of-the-art complementary metal-oxide-semiconductor (CMOS) logic, even accounting for the energy required to cryogenically cool the system \cite{gleb}.  However, existing superconducting logic families have limitations that complicate large-scale integration. For example, in rapid single flux quantum (RSFQ) logic \cite{RSFQ}, cumbersome splitter trees are required for gates with large fanout, precision lossy bias circuits are necessary for each gate, and high-density RAM remains elusive due the need for large inductors \cite{gleb}.  Other superconducting logic families, such as reciprocal quantum logic (RQL) \cite{RQL} and adiabatic quantum flux parametron (AQFP) \cite{AQFP} logic, require the distribution of AC clocking networks. These issues drastically limit throughput and scalability \cite{gleb}. 

Logic families based on signal levels instead of pulses enable lower-cost fan-out and RAM implementations and are an option to circumvent the above limitations.  A number of level-based superconducting technologies were proposed, including the cryotron \cite{cryotron}, which uses the stray magnetic field from a control wire to affect the superconductivity of a target superconductor.  To mitigate the large switching times of the cryotron, the tunneling cryotron \cite{tunnelingcryotron} replaced the target superconductor with a Josephson junction.  Similar gating devices including the nanocryotron (nTron) \cite{nTron} and heater-nanocryotron (hTron) \cite{hTron}, which gate superconductivity using current density and heat, respectively, incorporated analogous innovations.  All these devices have similar functionality -- suppressing the superconductivity of a target superconductor with the presence of a control current -- and therefore have similar weaknesses. First, these devices are only inverting, suppressing superconducting current when turned on, and non-inverting behavior is not possible, requiring lossy pull-up devices to achieve full swing of logic values.  Second, these devices are volatile, consuming significant static power.

We propose a novel non-volatile magnetic field-mediated superconducting (SuperMag) switch that can be constructed with either inverting or non-inverting behavior.  We experimentally demonstrate the central physical switching phenomenon, and propose a complete, directly cascadable logic family.  As no precision biasing or AC clocking is required, this logic family has the potential to outperform existing superconducting technologies for ultra-efficient computing.

\begin{figure}
    \centering
    \includegraphics[width=\linewidth]{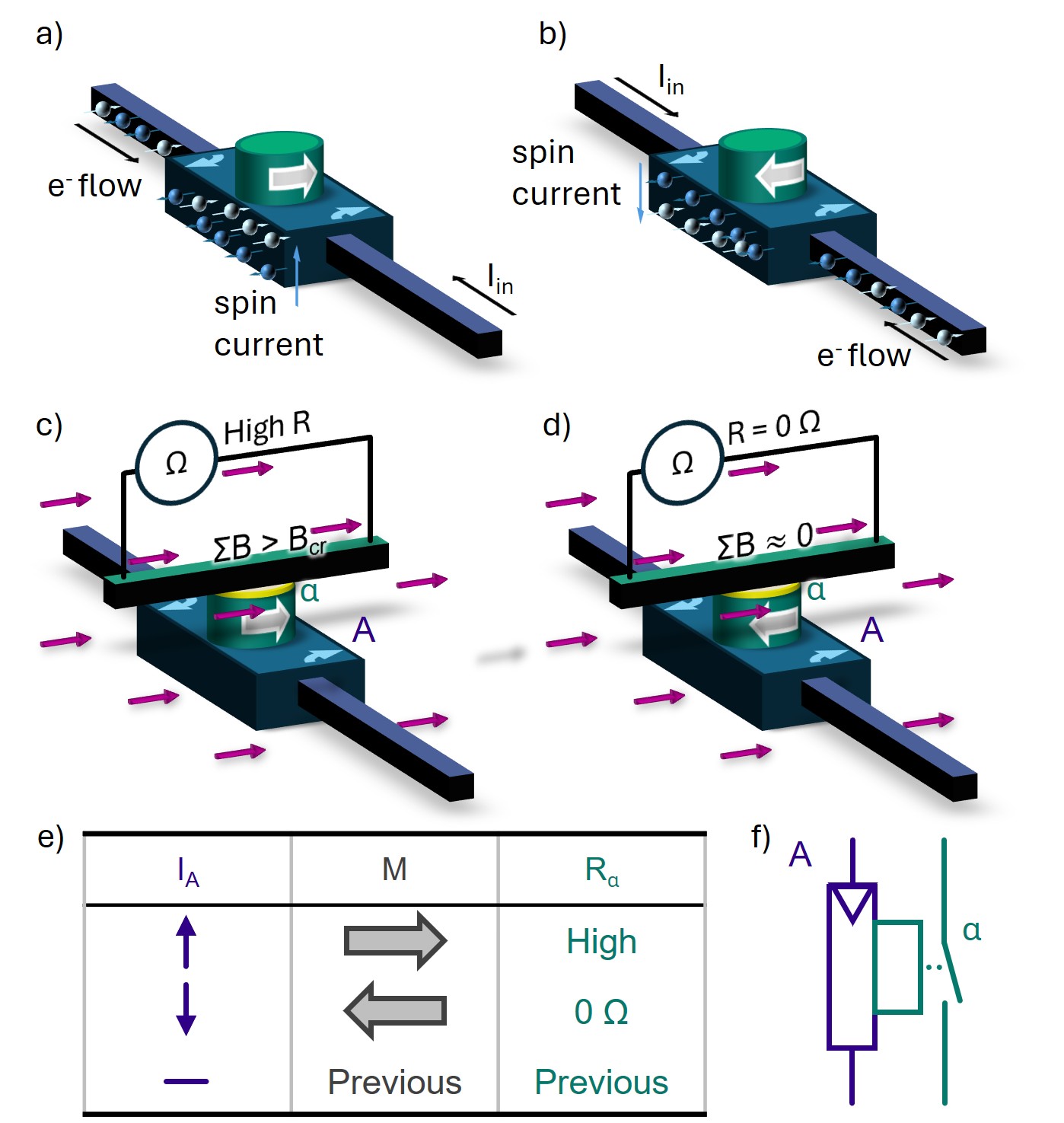}
    \caption{SuperMag switch. a) A current pulse through the SOT layer (blue rectangle), from bottom right to top left, switches the magnet's (green disk) magnetization to the right via the spin-Hall effect. b) Passing a current from top left to bottom right switches the magnetization to the left.  c) In the presence of a biasing field (purple arrows) and proximity coupled field from the magnet through an insulating layer (thin yellow disk), there is a sufficient magnetic field to cause the superconducting wire to be in a high resistance state.  d) When the proximity coupled field of the magnet opposes the bias field such that only a small field is felt by the superconductor, the superconductor is in a zero resistance state. e) Truth table for the SuperMag switch. f) SuperMag switch schematic symbol.  Current flowing in the same (opposite) direction of the triangle closes (opens) switch $\alpha$.}
     
    \label{fig:switch}
\end{figure}    

\begin{figure}
    \centering
    \includegraphics[width=\linewidth]{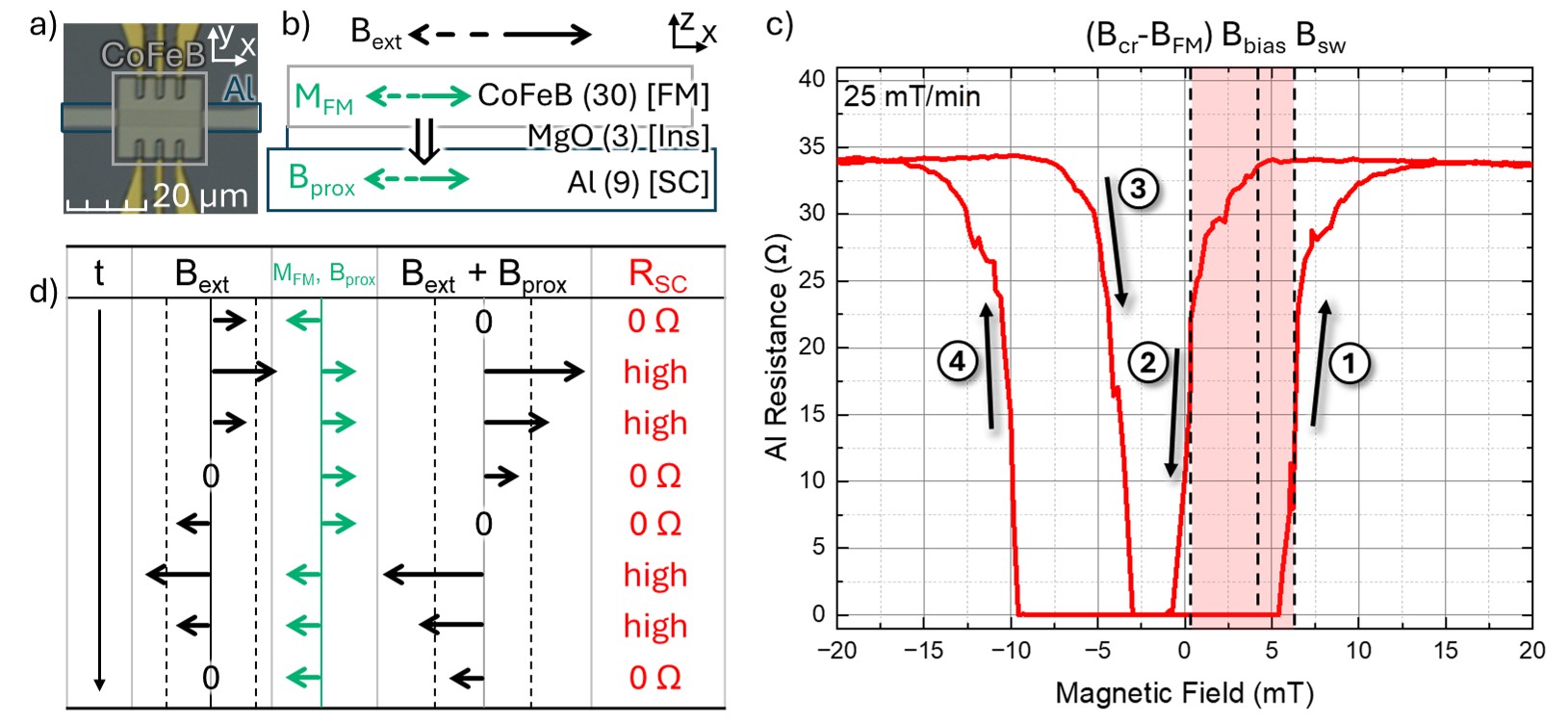}
    \caption{a) Fabricated Al/MgO/CoFeB device. b) Cross sectional illustration of device. B$_\tn{ext}$ is the in-plane external magnetic field, M$_\tn{FM}$ is the ferromagnet magnetization, B$_\tn{prox}$ is the effective field due to proximity effects felt by the superconductor.     c) (solid red) The experimentally measured resistance of the Al as a function of B$_\tn{ext}$ and M$_\tn{FM}$ at 270 mK. 
d) State variables of experiment through time. R$_\tn{SC}$ is the field-modulated resistance of the superconductor. Dashed lines in the B$_\tn{ext}$ (B$_\tn{ext} + \tn{B}_\tn{prox}$) column correspond with the switching threshold magnetic field of the ferromagnet (superconductor).}
    \label{fig:experimental}
\end{figure}

\section{Superconducting Magnetic Field Switch}

Depicted in Fig. \ref{fig:switch}, the SuperMag switch is constructed from a magnet proximity-coupled to a superconducting wire.  The magnet may be switched using spin-orbit torque (SOT) \cite{SOT, sot1, sot2, sot3, sot4} by passing a current through a heavy metal (HM) layer or topological insulator (TI) patterned beneath the magnet, as illustrated in Fig. \ref{fig:switch}a, b.  As depicted in Fig. \ref{fig:switch}c, d, the magnetization direction relative to a biasing magnetic field controls the resistance of the superconducting wire.  If the proximity-coupled field from the magnet (B$_{\tn{prox}}$) is in the same direction as the biasing field, $B_\tn{ext}$, the magnetic fields combine to surpass the critical magnetic field $B_{\tn{cr}}$, causing the superconductor to enter a high resistance state, and if B$_{\tn{prox}}$ opposes $B_\tn{ext}$, the fields cancel out, and the superconductor enters a zero resistance state.  Fig. \ref{fig:switch}e summarizes the switching behavior of the SuperMag switch.  The global biasing field, B$_\tn{ext}$, may be supplied by a superconducting magnet, patterned ferromagnets with fixed magnetization adjacent to each SuperMag switch (similar to the switches of \cite{2switch1, 2switch2, 2switch3, 2switch4}), large off-chip magnets, or engineered exchange bias in the FM.  

Passing a current through the SOT layer causes ferromagnetic switching according to the light blue curved arrows of Fig. \ref{fig:switch}a-d  -- the right-hand rule was arbitrarily chosen for the drawings in this work, but, in actuality, the chirality is material-dependent.  Superconductivity depends on the relative orientations of the two fields, and passing a current in one direction therefore causes the device to turn on (zero resistance), whereas passing a current in the other direction causes the device to turn off (high resistance).   Without loss of generality the chirality of the SOT switching can therefore be ignored, and a schematic symbol more simply encompassing the functionality of the SuperMag switch is illustrated in Fig. \ref{fig:switch}f. Current passed in the same (opposite) direction of the triangle point closes (opens) the switch, and the device retains its state after the current ceases to flow. 

\section{Experimental Validation}

We experimentally demonstrated the working principle of the SuperMag switch with an Al/MgO/CoFeB (SC/I/FM) heterostructure stack.  A micrograph and schematic cross section of our device is illustrated in Fig. \ref{fig:experimental}a,b.  The 9 nm single crystal superconducting Al film (SC) was first deposited on a Si(111) substrate using molecular beam epitaxy (MBE). The Al film was then patterned into a 5 $\mu$m wide channel using photolithography and wet etching. Photolithography was then used again to open an $\approx$10 $\mu$m x 20 $\mu$m area in the middle of the Al channel, followed by an e-beam deposition of a 3 nm MgO insulating layer (I) and a 30 nm CoFeB ferromagnetic layer (FM) to complete the test structure. Contacts were prefabricated on the substrate to electrically access the Al and the CoFeB. Crucially, the 3 nm MgO layer electrically insulates the CoFeB from the Al and is simultaneously thin enough to enable proximity coupling between the superconductor and the magnet.

The switching behavior of this heterostructure is depicted in Fig. \ref{fig:experimental}c, where the resistance of the Al channel was monitored as a function of in-plane applied magnetic field.  The measurements were performed at $^\tn{3}$He base temperature of $\approx$270 mK and with a bias current of 20 $\mu$A through the Al, which was chosen to produce the cleanest switching characteristics.   
The SC experiences the combined effect of both the external field, B$_\tn{ext}$, and a proximity exchange-coupled field from the FM \cite{prox_prb, switch_prl, switch_nmat} in the direction of magnetization of the FM, B$_\tn{prox} = \pm \tn{B}_\tn{FM}$, where B$_\tn{FM}$ is the magnitude of the proximity exchange-coupled field induced by the FM in the SC. If the absolute value of the combined external and exchange-coupled field experienced by the SC exceeds the critical field of the SC, \textit{i.e.} $|\tn{B}_{\tn{ext}} + \tn{B}_{\tn{prox}}| > \tn{B}_{\tn{cr}}$, then the SC will switch from superconducting to the normal resistive state.  Inversely, if the combined external and proximity field experienced by the SC is less than critical field of the SC, the SC will switch from normal to the superconducting state. 

The hysteretic switching behavior of the device is shown in Fig. \ref{fig:experimental}c,d. Initially, B$_\tn{ext} = 0$ T, and the magnetization of the FM points to the left, $(\tn{B}_\tn{prox} = -\tn{B}_\tn{FM})$.  Increasing B$_\tn{ext}$, eventually cancels the effect of B$_\tn{prox}$, experienced by the superconductor, $\tn{B}_\tn{ext} + (-\tn{B}_\tn{FM}) = 0$. When the magnitude of B$_\tn{ext}$ exceeds the coercive field of the FM, B$_\tn{sw}$, the magnetization of the FM is reversed to point in the direction of the applied field. B$_\tn{ext}$ and B$_\tn{prox}$ now positively combine, exceeding B$_\tn{cr}$, and turn the SC resistive, as illustrated in arm 1 of the switching curve. Increasing B$_\tn{ext}$ further keeps the SC in its normal resistive state. By decreasing B$_\tn{ext}$, the total field felt by the SC decreases and eventually the SC returns to its superconducting state when $\tn{B}_{\tn{ext}} + \tn{B}_{\tn{prox}} < \tn{B}_{\tn{cr}}$ (arm 2 of the switching curve). Sweeping B$_\tn{ext}$ over negative values produces similar switching characteristics.  Similar hysteretic switching behavior is observed in an identical material system with no MgO layer as depicted in Supp. Fig. \ref{suppfig:no-mgo}.

The device of Fig. \ref{fig:experimental}a requires minimal modification to function as a SuperMag switch. The gap between the two arms of the switching curve, \textit{i.e.} $\tn{B}_\tn{cr} - \tn{B}_\tn{FM} < \tn{B}_\tn{ext} < \tn{B}_\tn{sw}$, illustrated in pink in Fig. \ref{fig:experimental}c, is the foundation for utilizing this switching behavior for computing.  By applying a fixed bias magnetic field, B$_\tn{bias}$, within this critical range, the superconducting properties of the Al can be controlled solely by flipping the magnetization direction of the FM.  Additionally, the insulating behavior of the MgO layer enables complete electrical isolation between the FM and SC.  The FM can therefore be switched electrically without leaking current to the SC sensing wire, dramatically boosting energy efficiency.  Previously studied SC/FM heterostructures \cite{switch_prl, switch_nmat} show similar switching behavior but lack the input/output isolation demonstrated here, drastically reducing their attractiveness for logic.  In contrast, by adding an HM or TI layer atop the CoFeB and using SOT for switching \cite{SOT, sot1, sot2, sot3, sot4} the output current from one SuperMag device can be directly used as an input that switches another SuperMag device with minimal leakage.  Furthermore, as CoFeB and MgO are ubiquitous in commercial MRAM, Al/MgO/CoFeB is a strong candidate for creating SuperMag switches.

\section{SuperMag Logic and Memory}
\label{sec:logic-and-memory}

A full SuperMag processor -- logic and memory -- can be constructed from SuperMag switches.  As described throughout this section, SuperMag circuits are not limited to conventional CMOS circuit topologies, but rather, SuperMag offers numerous advantages that enable circuit-design-level optimizations in both logic and memory.  These advantages arise from a SuperMag switch's full-swing drive of both logic values, zero on-resistance, native signal inversion by swapping terminals, and non-volatility, and as none of these qualities are native to CMOS, SuperMag gates frequently have fewer-switch implementations compared with CMOS.

\subsection{SuperMag Logic}

Logic circuits are easily constructed from SuperMag switches.  While SuperMag switches have richer dynamics than conventional transistors, they can still be substituted one-to-one for transistors in CMOS logic circuits to achieve the same logical functionality.  This enables direct application of architectural, circuit design, and electronic design automation (EDA) techniques developed for CMOS.  Additionally, any logical function can be implemented in SuperMag using at most the same number of devices as in CMOS.  As described below, SuperMag is often more compact than CMOS, allowing for dense circuit designs.

\subsubsection{Inverter}

\begin{figure}
    \centering
    \includegraphics[width=\linewidth]{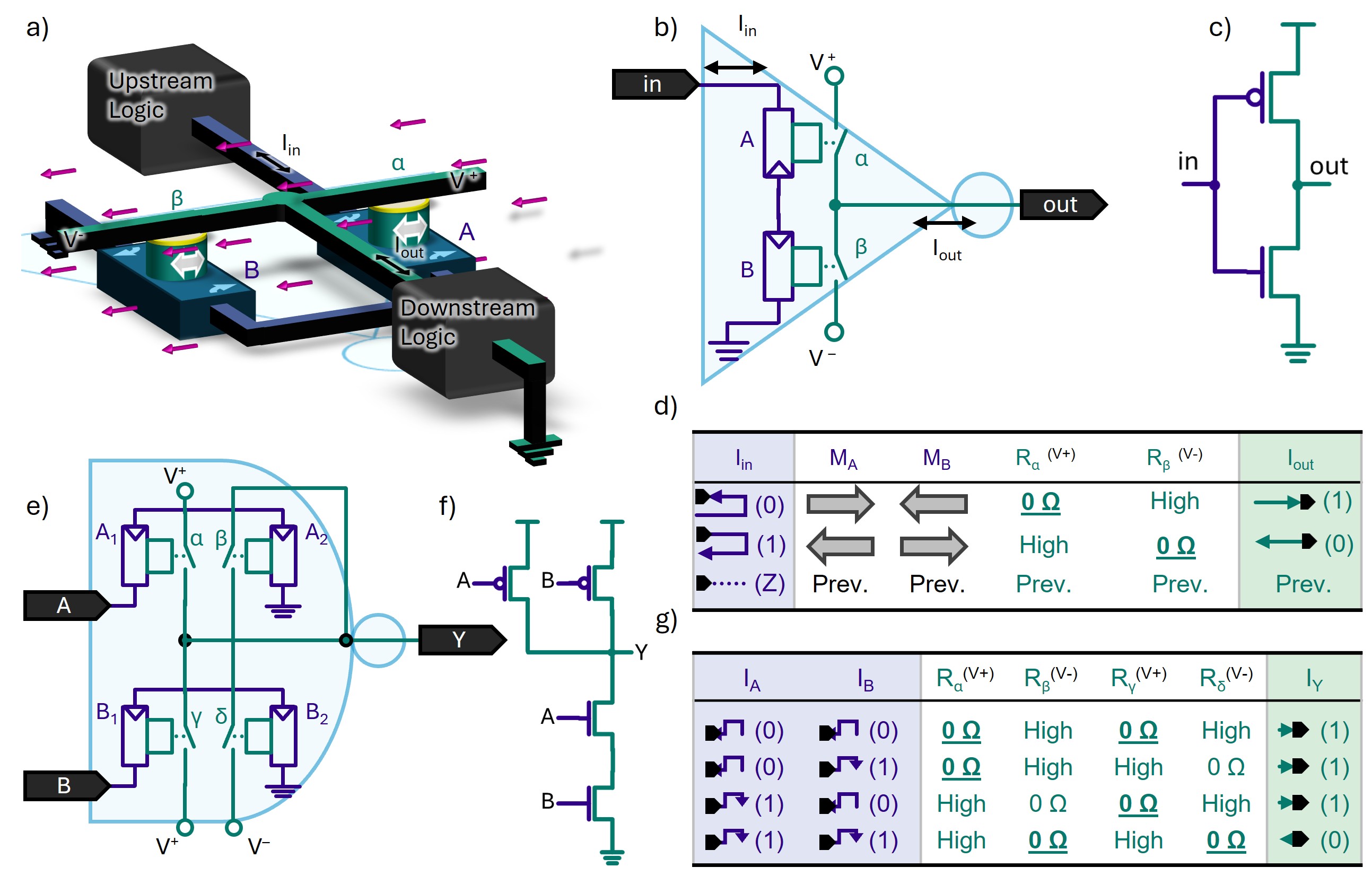}
    \caption{a-d) Complementary SuperMag inverter/buffer. a) 3D view.  The upstream logic drives the SOT layers of the SuperMag switches via the switching wire (blue).  The resistance of the output wire (green) is mediated by the magnetic state of the switches. b) SuperMag inverter schematic.  Throughout this work the color blue is used for circuit or switch inputs and switching wires and green for output wires and outputs in general.  A SuperMag buffer may be formed by swapping the $V^+$ and $V^-$ terminals.  c) CMOS inverter. d) SuperMag inverter truth table.  Bold, underlined ``0 $\Omega$''s indicate zero-resistance paths contributing to I$_\tn{out}$.  e-g) Complementary SuperMag NAND gate.  e) SuperMag NAND schematic. f) CMOS NAND gate. g) SuperMag NAND gate truth table.}
    \label{fig:inv}
\end{figure}

A SuperMag inverter can be constructed in three sections: switching wires (represented as blue in Fig. \ref{fig:inv}a, b), two SuperMag switches, and the output wire (represented as green).  Note the similarities in structure between the SuperMag switch and a CMOS inverter (Fig. \ref{fig:inv}c).  Input and output port direction is notated with the black arrow shapes.  A logic `1' (`0') is encoded by flowing current in the same (opposite) direction as the port arrow.  Therefore, current flowing into the inverter via the `in' port or out of the inverter via the `out' port represents a logic `1', and current in the opposite direction is a logic `0'.

Switching wire current flows between the `in' port and ground to switch the ferromagnets of the two SuperMag switches, thus modulating the superconductivity of the portions of the output wire coupled to the ferromagnets.  The two SuperMag switches are connected in a complementary manner and switch according to the triangles in Fig. \ref{fig:inv}b.  An input of `1' causes switch $\alpha$ to open (parallel fields) and switch $\beta$ to close (anti-parallel fields), creating a superconducting path in the output wire from V$^-$ through switch $\beta$ to the `out' port.  Inversely, if a logic `0' is presented to the `in' port, switch $\alpha$ closes and switch $\beta$ opens, causing the superconducting path to connect V$^+$ through switch $\alpha$ to `out'.

The `out' port further connects through the switching wires of downstream logic and eventually to ground.  Following an input to the inverter of logic `1' (`0'), this new superconducting path from V$^-$ (V$^+$) to `out' pulls (pushes) current into (out of) the inverter through the `out' port as a logic `0' (`1'). This completes the inverter truth table of Fig. \ref{fig:inv}d.

Additionally since SuperMag switches are non-volatile, the input does not need to be constantly driven once the inverter has computed the output.  Similar to \cite{FriedmanCMAT1}, sources V$^+$ and V$^-$ may be clocked to minimize static power dissipation.  This pulsing scheme also enables natural sequential pipelining and reduces the need for dedicated registers.

\subsubsection{NAND Gate}
\label{sec:nand}

A NAND gate is the smallest universal logic circuit, and multiple NAND gates can be connected to implement any logic function.  A SuperMag NAND gate, illustrated in Fig. \ref{fig:inv}e-g, is constructed from four SuperMag switches.  Each of the two inputs has its own switching wire driving two SuperMag switches.  As with the inverter, the switches on each switching wire are attached in a complementary manner such that exactly one switch is open (and one closed) after all switching events.  Presenting a logic `1' at input A (B) opens $\alpha$ ($\gamma$) and closes $\beta$ ($\delta$).  The V$^-$ output wire is connected in series through switches $\beta$ and $\delta$ necessitating that both inputs be `1' to produce a `0' at gate output `Y'.  In contrast, the output wire connected to $V^+$ is connected in parallel through $\alpha$ and $\gamma$, requiring only one input to be `0' for `Y' to produce a `1' output.  This completes the NAND gate truth table of Fig. \ref{fig:inv}g, and as all possible logic functions can be constructed using only NAND gates, SuperMag is therefore logically complete.

The SuperMag NAND gate of Fig. \ref{fig:inv}e and the CMOS NAND of Fig. \ref{fig:inv}f have similar structures.  As in CMOS, the output wire of the NAND is connected as a complementary network such that the output always has a low-resistance path to either to V$^+$ or V$^-$ but never both.  Likewise, any complementary logic gate that can be implemented in CMOS can be realized with a similar structure in SuperMag following the procedure outlined in Supplementary Note \ref{supnote:cmos-map}. Additionally, as further described in Supplementary Note \ref{supnote:cmos-map} and Supp. Fig. \ref{suppfig:swap}, due to the symmetric structure of the SuperMag switch, all inverting operations may be absorbed into upstream gate outputs or downstream gate inputs with zero overhead, greatly reducing the device count of SuperMag compared with CMOS needed to compute the same function.

\subsubsection{Transmission Gate Circuits}

SuperMag switches are perfect transmission gates: when turned on, there is a zero-resistance channel connecting both sides of the output wire.  This is in direct contrast to transistors, where the on-resistance of the source-drain path is non-negligible, necessitating frequent regeneration of signal strength, and strength depends on the driven logic value, necessitating two transistors for a good transmission gate.  As described in Supplementary Note \ref{supnote:tgates}, these perfect SuperMag transmission gates significantly reduce gate overhead compared to CMOS, and a single-SuperMag-switch tri-state buffer, which conventionally costs eight CMOS transistors, can be constructed.  Using similar techniques, a two-SuperMag-switch multiplexer (MUX) (normally comprising twelve CMOS transistors), and a six-SuperMag-switch exclusive-OR (XOR) gate (normally requiring at least ten CMOS transistors) can also be created.  

\subsubsection{Full Adder}

To illustrate the cascading and fanout of SuperMag logic gates, a SuperMag full adder is depicted in Fig. \ref{fig:f-adder}.  The adder comprises two XOR gates (as described in Supplementary Note \ref{supnote:tgates}) and an AO22 gate (i.e. 4-input AND-OR gate) constructed, as described in Supplementary Note \ref{supnote:cmos-map}, by mapping a CMOS AOI22 to SuperMag and swapping the V$^+$ and V$^-$ nodes to invert the output.

Cascading is accomplished by feeding the output of one gate to the input of the next.  Specifically, node X serves both as the output of the XOR gate in the upper left-hand side of the figure as well as an input to the other two logic gates.  This cascading of the output logic from one gate to the input logic of another gate is enabled by the fact that the gate inputs and outputs are represented by the same physical quantity: current direction. This enables seamless cascading without data converters or transformers between each gate.  

As SuperMag logic is current-based rather than voltage-based, fanout is accomplished by chaining the switching wires of the downstream logic in series (as opposed to in-parallel as in voltage-based CMOS).  This is depicted in the routing of the switching wires connected to A, B, Cin, and X in Fig. \ref{fig:f-adder}.  As current only needs to be as large as necessary to switch downstream magnets using SOT, current-induced Amperian magnetic fields of sensing wires of upstream switches -- being much less current-efficient than SOT for inducing ferromagnetic switching -- will not be large enough to disturb the non-volatile states of the upstream magnets. 

\begin{figure}
    \centering
    \includegraphics[width=\linewidth]{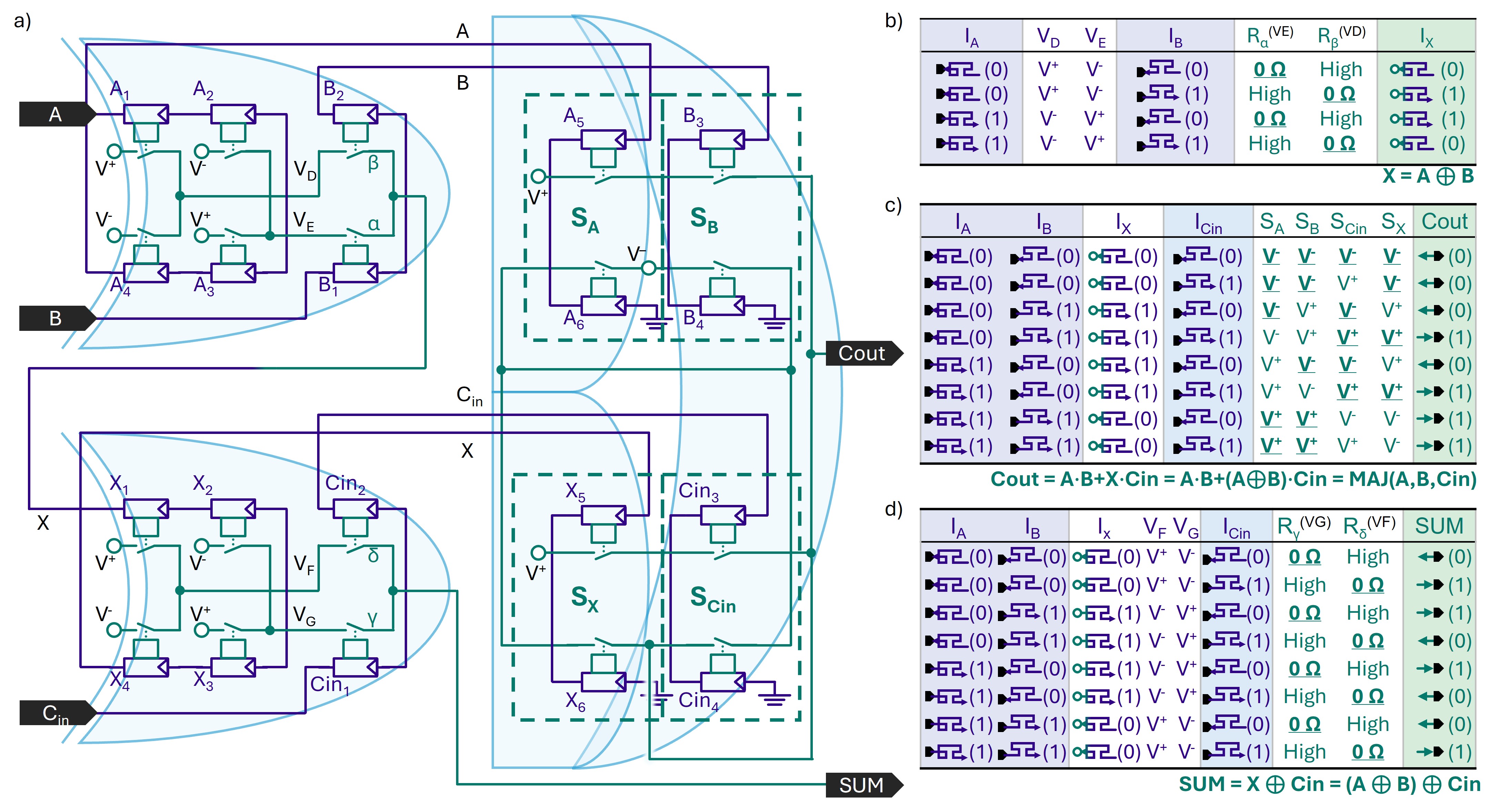}
    \caption{SuperMag one-bit full adder circuit structure and truth tables.  a) The full adder comprises two SuperMag XOR gates (left, similar to Supp. Fig. \ref{fig:xor}) and one AO22 gate (right). Cascade is accomplished by using the output of the upper-left XOR, X, to drive switching wires of both of the other gates.  Fanout is accomplished by chaining the SuperMag switches connected to A, B, and X in series.  b-d) Truth tables of the computation of X, SUM, and Cout respectively.  The S$_\tn{A}$, S$_\tn{B}$, S$_\tn{X}$, and S$_\tn{Cin}$ columns of (c) indicate the `on' branch of the corresponding complementary SuperMag switch pairs of (a).}
    \label{fig:f-adder}
\end{figure}

\subsection{SuperMag Memory}
\label{sec:aux}

The non-volatility of SuperMag combined with its perfect transmission gates enable memory functions that are far more compact than normally possible in CMOS.  As SuperMag switches are inherently stateful, they can be used to efficiently construct RAM.  This native RAM provides SuperMag with a distinct advantage over SFQ-based logic families where bit cells require SFQ$\rightarrow$DC current conversion \cite{gleb, rsfqbitcell} or heterogeneous technologies \cite{gleb}. SuperMag RAM has the added benefit of non-volatility even at room temperature, enabling long-term data storage and trivial power gating.  This is especially useful for retaining program or kernel storage in-between cryogenic cooling cycles, which is impossible when using superconducting current for memory storage. Along with native non-volatile RAM (nvRAM), SuperMag's non-volatility and transmission gates enable low-footprint implementations of other critical memory circuits such as latches, D flip-flops, and configuration storage for programmable logic as described in Supplementary Note \ref{supnote:secondary-memory}.  SuperMag's memory capabilities therefore make it very attractive for a wide spectrum of computing applications.

\begin{figure}
    \centering
    \includegraphics[width=\linewidth]{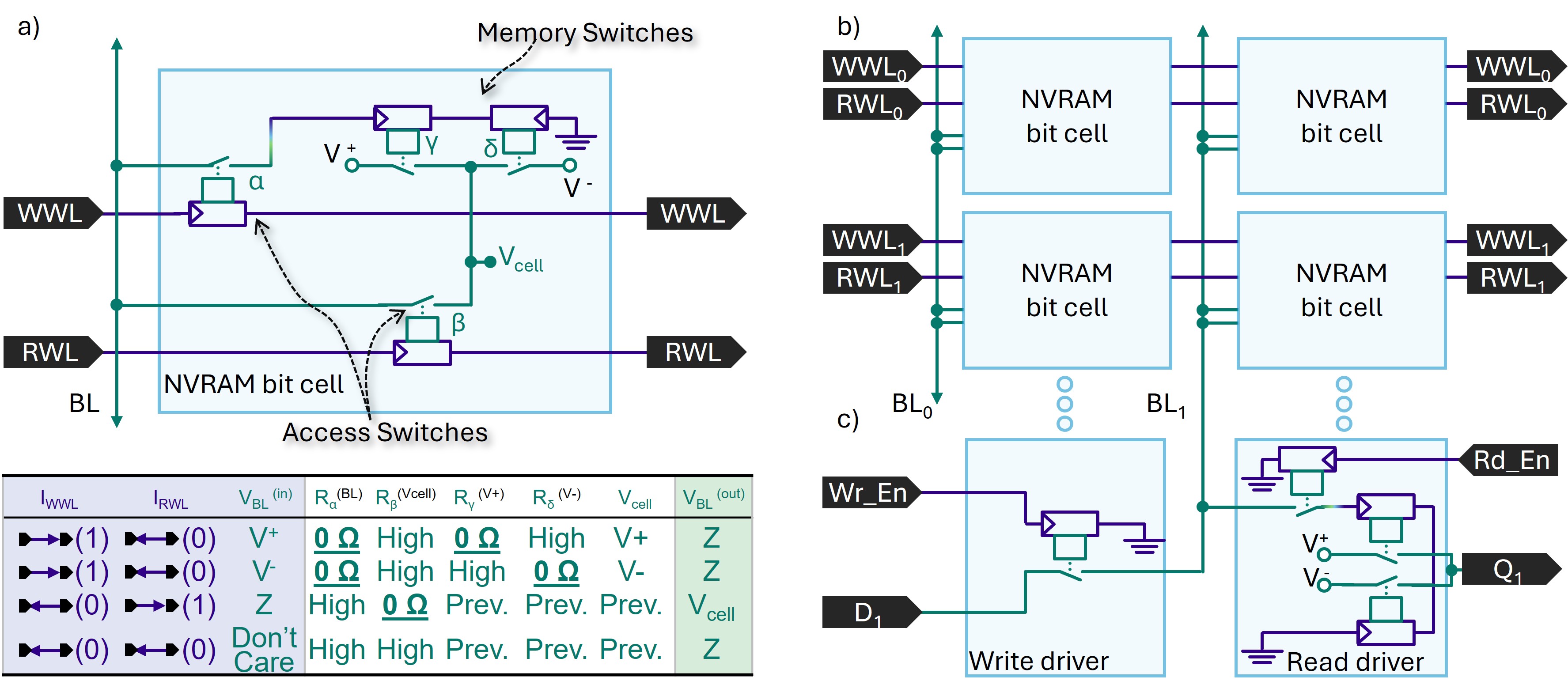}
    \caption{nvRAM with SuperMag. a) RAM bit cell and truth table.  Cell state is stored in SuperMag switches $\gamma$ and $\delta$, and read at node V$_\tn{cell}$.  If a `1' (`0') is stored, V$_\tn{cell}$ is driven to V$^+$ (V$^-$).  Writing is performed by driving WWL high and asserting BL with the bit to be written (V$^+$ or V$^-$).  The write access switch connects BL to the switching wires of switches $\delta$ and $\gamma$, overwriting the cell state.  To read, RWL is driven high, connecting V$_\tn{cell}$ to BL. b) nvRAM array comprising multiple bit cells.  c) Write and read drivers for column 1 of the nvRAM. Writing involves sending logic `1' to Wr\_En and activating the chosen WWL, followed by asserting D$_1$ with the data to be written.  Reading is performed by activating Rd\_En and one RWL to select a word.  The V$_\tn{cell}$ values of that word will drive the BLs and overwrite the the read driver buffers, which will latch the output data, thereby completing the read operation.}
    \label{fig:nvram}
\end{figure}

Fig. \ref{fig:nvram}a illustrates the construction of an nvRAM bit cell using SuperMag gates.  The bit cell comprises four SuperMag switches: the two access switches act as transmission gates, giving word line (WL) and bit line (BL) access to write or read from the cell, and two non-volatile memory switches store the bit of information.  To write to the cell, the corresponding write-WL (WWL) is driven with logic `1', and the BL is then driven with the bit of information to be stored.  Switch $\alpha$ closes, allowing the bit on the BL to overwrite the state of the SuperMag memory switches.  After writing, the WWL is driven low to disable writing of the cell, preventing it from being accidentally overwritten when the BL is used to read or write other cells in the column.

The bit cell state is stored in the memory switches and read via the read-WL (RWL).  If the cell stores a logic `1', $\gamma$ is closed, $\delta$ is open, and V$_\tn{cell}$ is driven with V$^+$; if the cell stores a logic `0', $\gamma$ is open, $\delta$ is closed, and V$_\tn{cell}$ is driven with V$^-$.
The cell state is read by sending a logic `1' through the RWL, enabling the read access switch.  When $\beta$ closes, V$_\tn{cell}$ is connected to the BL, allowing the cell state to be read.  After the cell has been read, the RWL is disabled to prevent contamination of the BL signal due to the cell state.

Fig. \ref{fig:nvram}b illustrates two words and two columns of a SuperMag nvRAM.  Each bit cell in a word is connected in series to the WWL and RWL for that word, and each column of bits is connected to that column's BL.  As in conventional CMOS RAM, all bit cells in a word are accessed simultaneously for both write and read operations.

Fig. \ref{fig:nvram}c illustrates the column write and read drivers for the SuperMag nvRAM.  BL access is controlled by the Wr\_En signal.  During a write, the chosen WWL is driven with `1' and Wr\_En with `1', allowing the D$_1$ data signal through the Write driver, BL, and cell access switch to overwrite the cell state.  After the cell has been written, WWL and Wr\_En are driven low.  

The Read driver block is used to conduct the read operation.  To initiate the read, the Rd\_En signal is driven high, and the desired RWL is enabled, connecting the BL to V$_\tn{cell}$ of the bit cell.  If the cell stores a logic `1' (`0'), BL is driven with V$^+$ (V$^-$), overwriting the Read buffer state with logic `1' (`0').  To complete the read, RWL and Rd\_En are driven with logic `0', and output Q$_1$ now reflects the state of the bit cell.

\section{Outlook}
\label{sec:outlook}

SuperMag has the potential to provide computational efficiency advantages when compared to other superconducting technologies and with CMOS.  These advantages can be seen through both a principal qualitative analysis of the unique physics of SuperMag and also, after anticipated optimization of SuperMag design and material properties, quantitatively via analysis with EDA software tools.  Additionally, Supplemetary Note \ref{supnote:sm_calcs} and Supplementary Tables \ref{tab:sc}-\ref{tab:func} highlight potential SuperMag material systems subject to SuperMag's functional constraints.

\subsection{Qualitative Advantages}

\begin{figure}
    \centering
    \includegraphics[width=\linewidth]{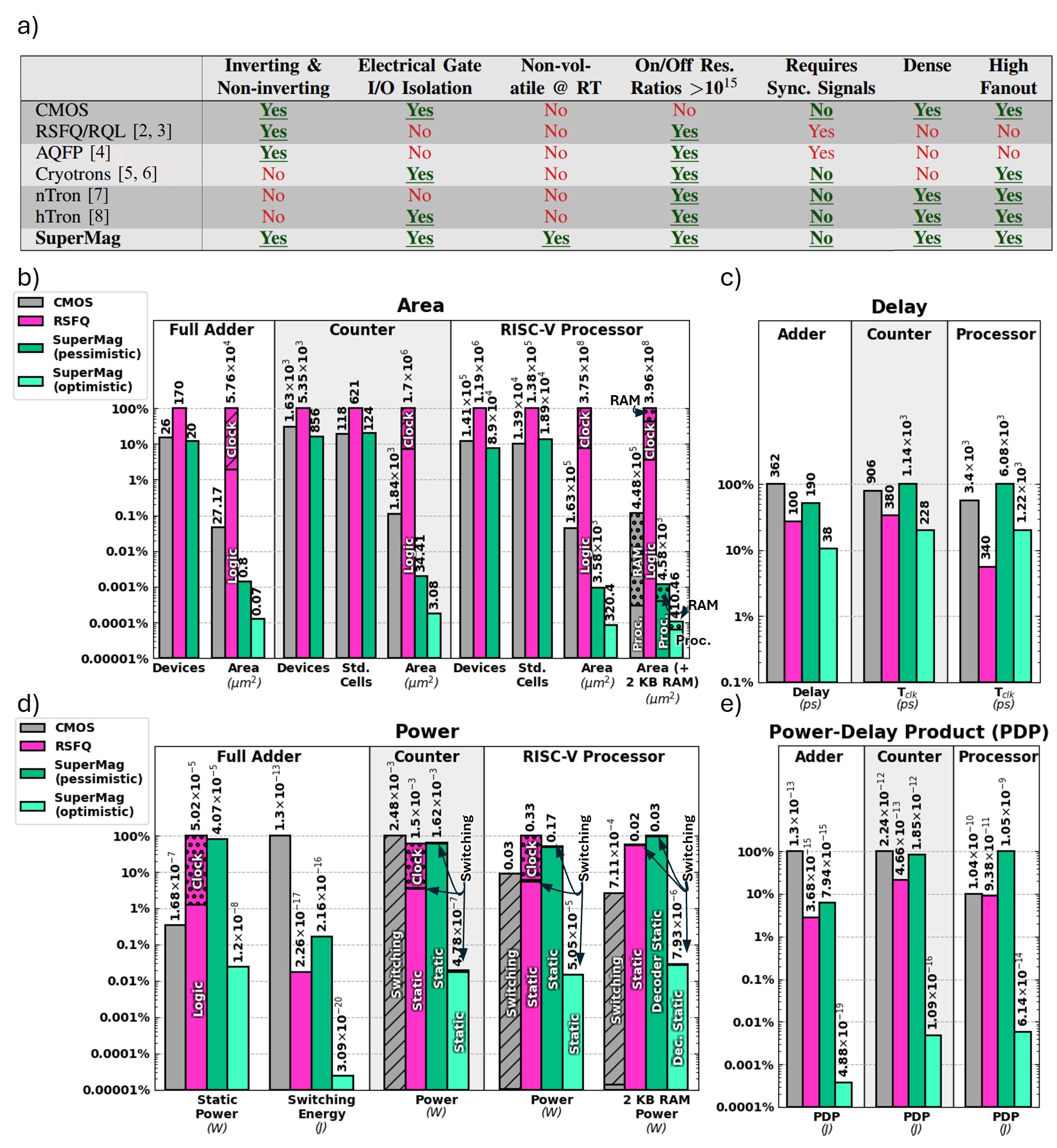}
    \caption{Comparison of SuperMag with other logic technologies.  a) Qualitative comparisons indicating SuperMag's potential as a replacement logic family.  b-e) Quantitative comparisons of area (b), delay (c), power (d), and power-delay-product PDP (e) evaluated for a combinational full adder, sequential 32-bit counter, and full RISC-V processor with and without an additional 2 KB RAM.  Tops of bars are normalized relative to the maximum value in that category and plotted on a logarithmic y-axis.  Some bars are partitioned into segments, the area of which are proportional to their contribution (not logarithmic).  e) PDP is calculated as product of power and delay and is therefore proportional to the energy required to complete the computation and inversely proportional the energy efficiency, often denoted with the unit, TOPS/W}
    \label{fig:comp}
\end{figure}

The table of Fig. \ref{fig:comp}a shows that the proposed SuperMag logic family has the potential to provide large advantages over CMOS and other superconducting logic families due to unique advantages inherent to the SuperMag switching mechanism.  SuperMag circumvents the need for lossy bias current distribution networks that plagues RSFQ, making SuperMag significantly easier to design and potentially more energy efficient.  Similarly, SuperMag does not require the precision AC clocking needed for RQL and AQFP, making SuperMag potentially faster than both.  Finally, while the cryotron, tunneling cryotron, nTron and hTron \cite{cryotron, tunnelingcryotron, nTron, hTron} utilize similar switching mechanisms to the SuperMag switch -- all five alter resistance of a superconductor -- the SuperMag switch has four distinct advantages:

\begin{itemize}[leftmargin=*]
    \item the SuperMag switch can be manufactured with inverting or non-inverting behavior, avoiding the need for dissipative pull-up devices and consuming significantly less energy;
    \item the two wires in the SuperMag switch are electrically isolated, greatly simplifying circuit complexity;
    \item the SuperMag switch is natively non-volatile, eliminating any static power needed to retain the logical state; and
    \item the SuperMag switch retains its state at room temperature, so programs, configurations, or processing states can be retained across cooling cycles, minimizing startup time. 
\end{itemize}

\noindent SuperMag thus takes advantage of the benefits provided by other superconducting logic families, without suffering from their drawbacks.

\subsection{Quantitative Advantages}

Based on analysis deatiled in Supplementary Note \ref{supnote:sm_calcs}, Fig. \ref{fig:comp}b-e illustrate the quantitative advantages of SuperMag when compared with CMOS (based on the SkyWater 130 nm process \cite{sky130}) and rapid SFQ (RSFQ) \cite{rsfqlib}.  Logic families were compared across three designs, a combinational full adder, a sequential 32-bit counter, and a full RISC-V processor \cite{ibex} with and without a 2 KB SRAM \cite{sky130ram, rsfqbitcell}.  Cadence genus and JoSIM \cite{josim}, were used for synthesis and simulation when necessary, and details of the computations are delineated in Supplementary Note \ref{supnote:comp}.

The SuperMag parameters listed in Supplementary Table \ref{tab:sm_params} were based on the material properties of NbN \cite{SOT} as the superconductor and Bi$_3$Sb$_2$/CoPt \cite{sot-bisb} as the SOT/FM system (see Supplementary Note \ref{supnote:sm_calcs}, Supplementary Tables \ref{tab:sc}-\ref{tab:func}, and \cite{sc_pb, sc_nb, SOT, sot4, sot-ptco, sot-bisb, sot2}).  Anticipated switching times and minimum spatial dimensions were gleaned from \cite{sot_switch_time}.  Additionally, a SuperMag based on more favorable material parameters (see Supplementary Note \ref{supnote:comp}) was also analyzed, labeled ``SuperMag (optimistic)''.

As illustrated in Fig. \ref{fig:comp}b, SuperMag is far more area efficient than both CMOS and RSFQ with a five order of magnitude (o.m.) reduction compared with RSFQ and a 1.5 o.m. reduction compared with CMOS.  Additionally, SuperMag uses roughly 37\% fewer devices than CMOS due to circuit design advantages outlined in Section \ref{sec:logic-and-memory}.

Fig. \ref{fig:comp}c illustrates the computational delay of the three technologies: RSFQ consistently dominates here due to no capacitive charging and no ferromagnetic switching.  SuperMag performs very similar to CMOS, and with future material advances, SuperMag computes more quickly for all designs, offering a potential two or three times speedup.

In Fig. \ref{fig:comp}d, power consumption is reported.  With today's material parameters, SuperMag performs very similarly to RSFQ with neglidgible switching power and with static power dominating dissipation.  Anticipating material optimizations, SuperMag consumes 100x less power than CMOS.  These computations do not include the power needed to cool the system, as significant cooling is not required in some environments, \textit{i.e.} space.  For situations where cooling is necessary, relative power comparisons of SuperMag and RSFQ do not change (as RSFQ also needs to be cooled), and SuperMag's efficiency boost compared to CMOS will be reduced.  In addition, as outlined in Supplementary Note \ref{supnote:sm_calcs}, total power consumption of SuperMag scales very favorably, being proportional to the volume of the switch which ideally scales down cubically with feature size.

Finally, Fig. \ref{fig:comp}e lists the power-delay product (PDP) which is proportional to the average energy required for a computation and inversely proportional to the computational energy, often reported in TOPS/W.  With optimized material parameters, SuperMag performs very well, with materials only needing to be improved by $\approx$ 32\% (see Supplementary Note \ref{supnote:comp}) before SuperMag becomes competitive with mature technologies.  With anticipated full optimizations, SuperMag provides a 1000x boost in computational efficiency compared with the mature technologies.

SuperMag therefore offers large potential energy savings, though the present constraints are tight.  Incorporating optimized material engineering and the architectural benefits introduced in Section \ref{sec:logic-and-memory} can unlock these energy savings, ushering in a new generation of ultra-efficient digital logic.

\bibliography{main}

@ARTICLE{FriedmanCMAT1,
  author={Friedman, Joseph S. and Sahakian, Alan V.},
  journal={IEEE TED}, 
  title={Complementary Magnetic Tunnel Junction Logic}, 
  year={2014},
  volume={61},
  number={4},
  pages={1207-1210},
  doi={10.1109/TED.2014.2306395}}

@article{hTron,
  title = {Multilayered Heater Nanocryotron: A Superconducting-Nanowire-Based Thermal Switch},
  author = {Baghdadi, Reza and Allmaras, Jason P. and Butters, Brenden A. and Dane, Andrew E. and Iqbal, Saleem and McCaughan, Adam N. and Toomey, Emily A. and Zhao, Qing-Yuan and Kozorezov, Alexander G. and Berggren, Karl K.},
  journal = {Phys. Rev. Appl.},
  volume = {14},
  issue = {5},
  pages = {054011},
  numpages = {12},
  year = {2020},
  month = {Nov},
  publisher = {American Physical Society},
  doi = {10.1103/PhysRevApplied.14.054011},
}

@Article{nTron,
author={McCaughan, Adam N.
and Berggren, Karl K.},
title={A Superconducting-Nanowire Three-Terminal Electrothermal Device},
journal={Nano Letters},
year={2014},
month={Oct},
day={08},
publisher={American Chemical Society},
volume={14},
number={10},
pages={5748-5753},
issn={1530-6984},
doi={10.1021/nl502629x},
}

@Article{SOT,
author={Nguyen, Minh-Hai and Ribeill, Guilhem J. and Gustafsson, Martin V. and Shi, Shengjie and Aradhya, Sriharsha V. and Wagner, Andrew P. and Ranzani, Leonardo M. and Zhu, Lijun and Baghdadi, Reza and Butters, Brenden and Toomey, Emily and Colangelo, Marco and Truitt, Patrick A. and Jafari-Salim, Amir and McAllister, David and Yohannes, Daniel and Cheng, Sean R. and Lazarus, Rich and Mukhanov, Oleg and Berggren, Karl K. and Buhrman, Robert A. and Rowlands, Graham E. and Ohki, Thomas A.},
title={Cryogenic Memory Architecture Integrating Spin Hall Effect based Magnetic Memory and Superconductive Cryotron Devices},
journal={Scientific Reports},
year={2020},
month={Jan},
day={14},
volume={10},
number={1},
pages={248},
issn={2045-2322},
doi={10.1038/s41598-019-57137-9},
}

@book{gleb,
    TITLE = {Single Flux Quantum Integrated Circuit Design},
  AUTHOR = {Krylov, Gleb and Friedman, Eby G.},
  YEAR = {2021},
  PUBLISHER = {Springer},
}

@article{RQL,
    author = {Herr, Quentin P. and Herr, Anna Y. and Oberg, Oliver T. and Ioannidis, Alexander G.},
    title = {Ultra-low-power superconductor logic},
    journal = {Journal of Applied Physics},
    volume = {109},
    number = {10},
    pages = {103903},
    year = {2011},
    month = {05},
    issn = {0021-8979},
    doi = {10.1063/1.3585849},
    url = {https://doi.org/10.1063/1.3585849},
    eprint = {https://pubs.aip.org/aip/jap/article-pdf/doi/10.1063/1.3585849/15080859/103903\_1\_online.pdf},
}

@ARTICLE{AQFP,
  author={Harada, Y. and Nakane, H. and Miyamoto, N. and Kawabe, U. and Goto, E. and Soma, T.},
  journal={IEEE Transactions on Magnetics}, 
  title={Basic operations of the quantum flux parametron}, 
  year={1987},
  volume={23},
  number={5},
  pages={3801-3807},
  doi={10.1109/TMAG.1987.1065574}}

@ARTICLE{RSFQ,
  author={Likharev, K.K. and Semenov, V.K.},
  journal={IEEE Transactions on Applied Superconductivity}, 
  title={RSFQ logic/memory family: a new Josephson-junction technology for sub-terahertz-clock-frequency digital systems}, 
  year={1991},
  volume={1},
  number={1},
  pages={3-28},
  doi={10.1109/77.80745}}

@ARTICLE{cryotron,
  author={Buck, D. A.},
  journal={Proceedings of the IRE}, 
  title={The Cryotron-A Superconductive Computer Component}, 
  year={1956},
  volume={44},
  number={4},
  pages={482-493},
  doi={10.1109/JRPROC.1956.274927}}

@ARTICLE{tunnelingcryotron,
  author={Matisoo, J.},
  journal={Proceedings of the IEEE}, 
  title={The tunneling cryotron: A superconductive logic element based on electron tunneling}, 
  year={1967},
  volume={55},
  number={2},
  pages={172-180},
  doi={10.1109/PROC.1967.5436}}

@article{sot1,
  title = {Fieldlike and antidamping spin-orbit torques in as-grown and annealed Ta/CoFeB/MgO layers},
  author = {Avci, Can Onur and Garello, Kevin and Nistor, Corneliu and Godey, Sylvie and Ballesteros, Bel\'en and Mugarza, Aitor and Barla, Alessandro and Valvidares, Manuel and Pellegrin, Eric and Ghosh, Abhijit and Miron, Ioan Mihai and Boulle, Olivier and Auffret, Stephane and Gaudin, Gilles and Gambardella, Pietro},
  journal = {Phys. Rev. B},
  volume = {89},
  issue = {21},
  pages = {214419},
  numpages = {13},
  year = {2014},
  month = {Jun},
  publisher = {American Physical Society},
  doi = {10.1103/PhysRevB.89.214419},
  url = {https://link.aps.org/doi/10.1103/PhysRevB.89.214419}
}

@article{sot2,
    author = {Zhang, C. and Fukami, S. and Sato, H. and Matsukura, F. and Ohno, H.},
    title = {Spin-orbit torque induced magnetization switching in nano-scale Ta/CoFeB/MgO},
    journal = {Applied Physics Letters},
    volume = {107},
    number = {1},
    pages = {012401},
    year = {2015},
    month = {07},
    issn = {0003-6951},
    doi = {10.1063/1.4926371},
    url = {https://doi.org/10.1063/1.4926371},
    eprint = {https://pubs.aip.org/aip/apl/article-pdf/doi/10.1063/1.4926371/13968659/012401\_1\_online.pdf},
}

@article{sot3,
    author = {Takeuchi, Yutaro and Zhang, Chaoliang and Okada, Atsushi and Sato, Hideo and Fukami, Shunsuke and Ohno, Hideo},
    title = {Spin-orbit torques in high-resistivity-W/CoFeB/MgO},
    journal = {Applied Physics Letters},
    volume = {112},
    number = {19},
    pages = {192408},
    year = {2018},
    month = {05},
    issn = {0003-6951},
    doi = {10.1063/1.5027855},
    url = {https://doi.org/10.1063/1.5027855},
    eprint = {https://pubs.aip.org/aip/apl/article-pdf/doi/10.1063/1.5027855/14141386/192408\_1\_online.pdf},
}

@article{sot4,
    author = {Lee, Hae-Yeon and Kim, Sanghoon and Park, June-Young and Oh, Young-Wan and Park, Seung-Young and Ham, Wooseung and Kotani, Yoshinori and Nakamura, Tetsuya and Suzuki, Motohiro and Ono, Teruo and Lee, Kyung-Jin and Park, Byong-Guk},
    title = {Enhanced spin-orbit torque via interface engineering in Pt/CoFeB/MgO heterostructures},
    journal = {APL Materials},
    volume = {7},
    number = {3},
    pages = {031110},
    year = {2019},
    month = {03},
    issn = {2166-532X},
    doi = {10.1063/1.5084201},
    url = {https://doi.org/10.1063/1.5084201},
    eprint = {https://pubs.aip.org/aip/apm/article-pdf/doi/10.1063/1.5084201/14561476/031110\_1\_online.pdf},
}

@article{prox_prb,
  title = {Inhomogeneous magnetism induced in a superconductor at a superconductor-ferromagnet interface},
  author = {Krivoruchko, V. N. and Koshina, E. A.},
  journal = {Phys. Rev. B},
  volume = {66},
  issue = {1},
  pages = {014521},
  numpages = {6},
  year = {2002},
  month = {Jul},
  publisher = {American Physical Society},
  doi = {10.1103/PhysRevB.66.014521},
  url = {https://link.aps.org/doi/10.1103/PhysRevB.66.014521}
}

@article{switch_prl,
  title = {Superconducting Spin Switch with Infinite Magnetoresistance Induced by an Internal Exchange Field},
  author = {Li, Bin and Roschewsky, Niklas and Assaf, Badih A. and Eich, Marius and Epstein-Martin, Marguerite and Heiman, Don and M\"unzenberg, Markus and Moodera, Jagadeesh S.},
  journal = {Phys. Rev. Lett.},
  volume = {110},
  issue = {9},
  pages = {097001},
  numpages = {5},
  year = {2013},
  month = {Feb},
  publisher = {American Physical Society},
  doi = {10.1103/PhysRevLett.110.097001},
  url = {https://link.aps.org/doi/10.1103/PhysRevLett.110.097001}
}

@Article{switch_nmat,
author={Zhu, Yi
and Pal, Avradeep
and Blamire, Mark G.
and Barber, Zoe H.},
title={Superconducting exchange coupling between ferromagnets},
journal={Nature Materials},
year={2017},
month={Feb},
day={01},
volume={16},
number={2},
pages={195-199},
issn={1476-4660},
doi={10.1038/nmat4753},
url={https://doi.org/10.1038/nmat4753}
}

@article{sot-ptco,
author={Xue, Fen and Lin, Shy-Jay and Song, Mingyuan and Hwang, William and Klewe, Christoph and Lee, Chien-Min and Turgut, Emrah and Shafer, Padraic and Vailionis, Arturas and Huang, Yen-Lin and Tsai, Wilman and Bao, Xinyu and Wang, Shan X.},
title={Field-free spin-orbit torque switching assisted by in-plane unconventional spin torque in ultrathin [Pt/Co]N},
journal={Nature Communications},
year={2023},
month={Jul},
day={04},
volume={14},
number={1},
pages={3932},
issn={2041-1723},
doi={10.1038/s41467-023-39649-1},
}

@Article{sot-bisb,
author={Fan, Tuo and Khang, Nguyen Huynh Duy and Nakano, Soichiro and Hai, Pham Nam},
title={Ultrahigh efficient spin orbit torque magnetization switching in fully sputtered topological insulator and ferromagnet multilayers},
journal={Scientific Reports},
year={2022},
month={Feb},
day={22},
volume={12},
number={1},
pages={2998},
issn={2045-2322},
doi={10.1038/s41598-022-06779-3},
}

@ARTICLE{sc_nb,
  author={Huebener, R. and Kampwirth, R. and Martin, R. and Barbee, T. and Zubeck, R.},
  journal={IEEE Transactions on Magnetics}, 
  title={Critical current density in superconducting niobium films}, 
  year={1975},
  volume={11},
  number={2},
  pages={344-346},
  doi={10.1109/TMAG.1975.1058718}}

@article{sc_pb,
  title = {Robust superconductivity in quantum-confined Pb: Equilibrium and irreversible superconductive properties},
  author = {\"Ozer, Mustafa M. and Thompson, James R. and Weitering, Hanno H.},
  journal = {Phys. Rev. B},
  volume = {74},
  issue = {23},
  pages = {235427},
  numpages = {11},
  year = {2006},
  month = {Dec},
  publisher = {American Physical Society},
  doi = {10.1103/PhysRevB.74.235427},
  url = {https://link.aps.org/doi/10.1103/PhysRevB.74.235427}
}

@article{2switch2,
  title = {Low-Field Superconducting Spin Switch Based on a Superconductor $/$Ferromagnet Multilayer},
  author = {Tagirov, L. R.},
  journal = {Phys. Rev. Lett.},
  volume = {83},
  issue = {10},
  pages = {2058--2061},
  numpages = {0},
  year = {1999},
  month = {Sep},
  publisher = {American Physical Society},
  doi = {10.1103/PhysRevLett.83.2058},
  url = {https://link.aps.org/doi/10.1103/PhysRevLett.83.2058}
}

@article{2switch3,
  title = {Ferromagnetic-semiconductor--singlet-(or triplet) superconductor--ferromagnetic-semiconductor systems as possible logic circuits and switches},
  author = {Kuli\ifmmode \acute{c}\else \'{c}\fi{}, Miodrag L. and Endres, Martin},
  journal = {Phys. Rev. B},
  volume = {62},
  issue = {17},
  pages = {11846--11853},
  numpages = {0},
  year = {2000},
  month = {Nov},
  publisher = {American Physical Society},
  doi = {10.1103/PhysRevB.62.11846},
  url = {https://link.aps.org/doi/10.1103/PhysRevB.62.11846}
}

@article{2switch1,
title = {Coupling between ferromagnets through a superconducting layer},
journal = {Physics Letters},
volume = {23},
number = {1},
pages = {10-11},
year = {1966},
issn = {0031-9163},
doi = {https://doi.org/10.1016/0031-9163(66)90229-0},
url = {https://www.sciencedirect.com/science/article/pii/0031916366902290},
author = {P.G. {De Gennes}},
}

@Article{2switch4,
author={Matsuki, Hisakazu and Hijano, Alberto and Mazur, Grzegorz P.  and Ili{\'{c}}, Stefan and Wang, Binbin and Alekhina, Iuliia and Ohnishi, Kohei and Komori, Sachio and Li, Yang and Stelmashenko, Nadia and Banerjee, Niladri and Cohen, Lesley F.  and McComb, David W.  and Bergeret, F. Sebasti{\'a}n and Yang, Guang and Robinson, Jason W. A.},
title={Realisation of de Gennes' absolute superconducting switch with a heavy metal interface},
journal={Nature Communications},
year={2025},
month={Jul},
day={01},
volume={16},
number={1},
pages={5674},
issn={2041-1723},
doi={10.1038/s41467-025-61267-2},
url={https://doi.org/10.1038/s41467-025-61267-2}
}

@misc{sky130,
  author       = {Google LLC and SkyWater Technology},
  title        = {SkyWater Open Source PDK},
  year         = {2020},
  howpublished = {\url{https://github.com/google/skywater-pdk}},
  note         = {Accessed 2025-11-11}
}

@misc{sky130ram,
  author       = {UC Santa Cruz VLSI Design and Automation research lab},
  title        = {SKY130 Standard SRAM configurations},
  year         = {2024},
  howpublished = {\url{https://github.com/VLSIDA/sky130_sram_macros}},
  note         = {Accessed 2025-11-11}
}

@misc{rsfqlib,
  author       = {Schindler, Lieze and Hall, Tessa},
  title        = {ColdFlux RSFQ Logic Cell Library for MIT-LL SFQ Process},
  year         = {2023},
  howpublished = {\url{https://github.com/sunmagnetics/RSFQlib}},
  note         = {Version 3.0, accessed 2025-11-11}
}

@misc{josim,
  author       = {Delport, Johannes A.},
  title        = {JoSIM: Superconductor Circuit Simulator},
  year         = {2023},
  howpublished = {\url{https://github.com/JoeyDelp/JoSIM}},
  note         = {Version 2.6.8, accessed 2025-11-11}
}

@ARTICLE{rsfqbitcell,
  author={Semenov, Vasili K. and Polyakov, Yuri A. and Tolpygo, Sergey K.},
  journal={IEEE Transactions on Applied Superconductivity}, 
  title={Very Large Scale Integration of Josephson-Junction-Based Superconductor Random Access Memories}, 
  year={2019},
  volume={29},
  number={5},
  pages={1-9},
  doi={10.1109/TASC.2019.2904971}
}

@misc{ibex,
  author       = {lowRISC C.I.C.},
  title        = {Ibex RISC-V Core},
  year         = {2025},
  howpublished = {\url{https://github.com/lowRISC/ibex}},
  note         = {Accessed 2025-11-11}
}

@article{sot_switch_time,
    author = {Nguyen, T. V. A. and Naganuma, H. and Honjo, H. and Ikeda, S. and Endoh, T.},
    title = {Ultrafast spin-orbit torque-induced magnetization switching in a 75°-canted magnetic tunnel junction},
    journal = {AIP Advances},
    volume = {14},
    number = {2},
    pages = {025018},
    year = {2024},
    month = {02},
    issn = {2158-3226},
    doi = {10.1063/9.0000789},
    url = {https://doi.org/10.1063/9.0000789}
}

\vfill
\pagebreak

%
%

\setcounter{page}{1}

\setcounter{section}{0}
\renewcommand{\thesection}{S\arabic{section}}
\titleformat{\section}{\Large\filcenter\bfseries}{Supplementary Note \thesection:}{5pt}{}[]
\titleformat{\subsection}{\Large\itshape\filright}{Supplementary Note \thesection:}{5pt}{}

\onecolumn
\setcounter{figure}{0}
\renewcommand{\figurename}{Supp. Fig.}
\renewcommand{\thefigure}{S\arabic{figure}}

\setcounter{table}{0}
\renewcommand{\tablename}{SUPPLEMENTARY TABLE}
\renewcommand{\thetable}{S\Roman{table}}

\setcounter{equation}{0}
\renewcommand{\theequation}{s-\arabic{equation}}

\centerline{\LARGE{Supplementary Information for}}
\centerline{\LARGE{``Magnetic Field-Mediated Superconducting Logic''}}

\vspace{1em}

\centerline{\large{Alexander J. Edwards*$^{1,2}$, Son T. Le*$^1$, Nicholas W. G. Smith$^1$, Ebenezer C. Usih$^2$,}}

\centerline{\large{Austin Thomas $^1$, Christopher J. K. Richardson$^1$, Nicholas A. Blumenschein$^1$}}

\centerline{\large{Aubrey T. Hanbicki$^1$, Adam L. Friedman$^1$, Joseph S. Friedman$^2$}}

\vspace{1em}

\centerline{\large{$^1$Laboratory for Physical Sciences, College Park, MD, USA}}
\vspace{0.5em}
\centerline{\large{$^2$Department of Electrical and Computer Engineering}}

\centerline{\large{The University of Texas at Dallas, Richardson, TX, USA}}

\pagebreak

\begin{figure}[h]
    \centering
    \includegraphics[width=0.9\linewidth]{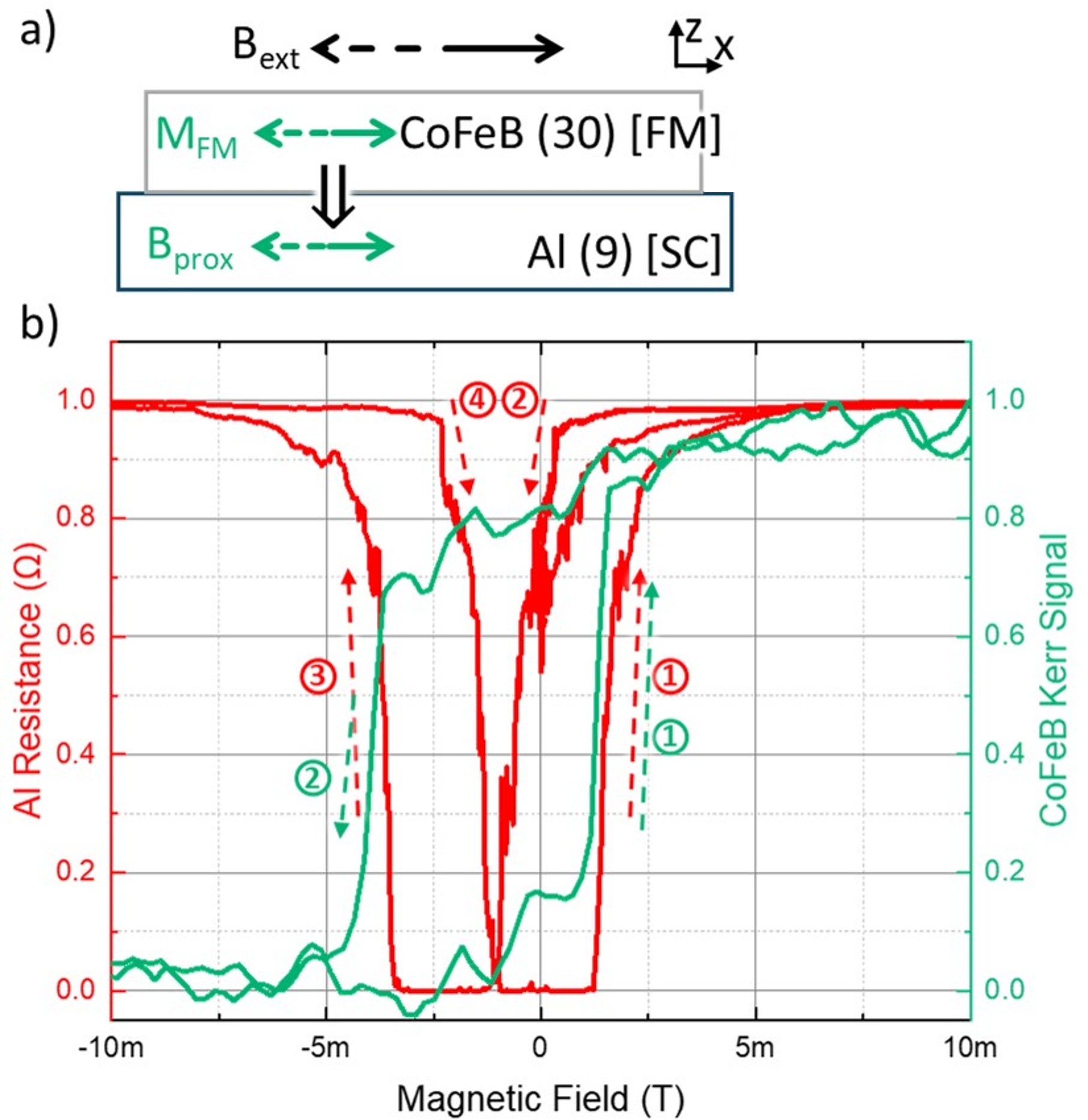}
    \caption{Switching characteristics of an Al/CoFeB heterostructure stack, a separate device from the device shown in the main text. a) Device illustration. The device’s fabrication is similar to the Al/MgO/CoFeB stack used in the main text but without the MgO insulating layer deposition step. Here, the Al and CoFeB are in direct interface with each other to enhance the proximity exchange coupling effect and confirm the hysteresis switching behavior of the Al channel’s resistance as function of external field. Due to the lack of input/output insulation, this device structure has limited application in scaled logic. b) Low-temperature ($\approx$270 mK) transport measurement (Red) showing the switching characteristic of the Al (SC) as function of applied in-plane magnetic field, and Magnetic Optical Kerr Effect (MOKE) measurement (Green) of the CoFeB layer at 10 K showing the magnetization properties of the ferromagnet. Interestingly, the CoFeB switches at approximately the same field as the superconductor indicating that the superconductivity of the SC is induced by the switching of the FM.}
    \label{suppfig:no-mgo}    
\end{figure}

\pagebreak

\section{\\CMOS-Mapped Logic Gates}
\label{supnote:cmos-map}

Any complementary logic gate that can be implemented in CMOS can be realized with a similar structure in SuperMag.  As in CMOS, \textit{each} logical input is connected to a complementary pair of SuperMag switches.  The routing of the output wire is identical to the transistor connectivity in CMOS.  This mapping is seen in the NAND gate of Fig. \ref{fig:inv}e, f where the pull-up PMOS transistors in the CMOS NAND gate connect V$_\tn{DD}$ to the output in parallel.  Thus, in the SuperMag NAND, switches $\alpha$ and $\gamma$ are connected in parallel.  Likewise, the NMOS transistors in a CMOS gate are connected in series, so the output wire of the SuperMag gate is connected in series through switches $\beta$ and $\delta$.  This mapping may be similarly applied to any complementary CMOS gate.

\begin{figure}[ht]
    \centering
    \includegraphics[width=0.9\linewidth]{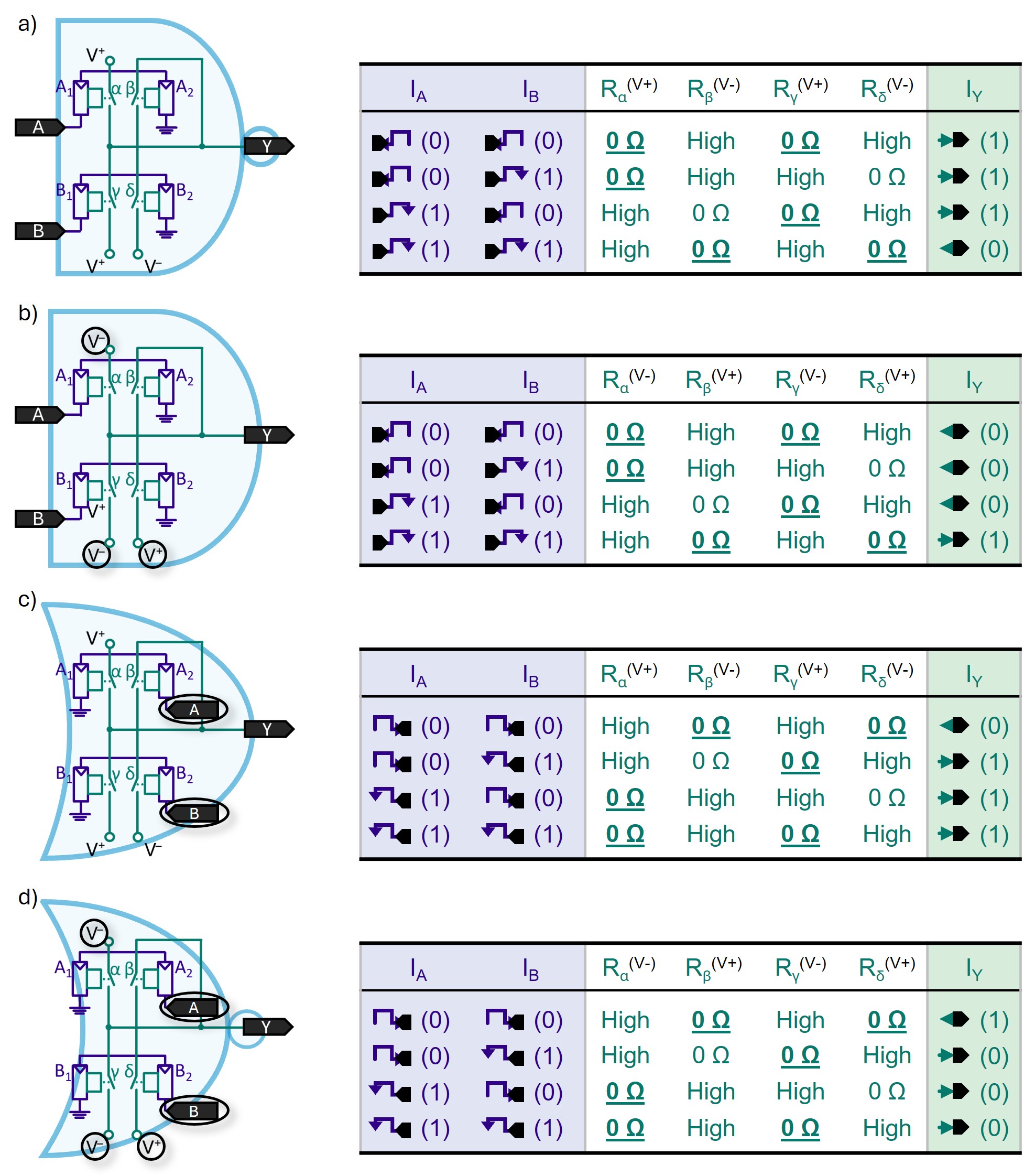}
    \caption{SuperMag NAND (a), AND (b), OR (c), and NOR (d) gates and truth tables.  Switch layout and routing are identical for all four circuits.  Polymorphism is achieved simply by swapping V$^+$ and V$^-$ (inverting the output) and swapping the input current direction (inverting the inputs).  Neither of these polymorphism techniques are possible in CMOS.}
    \label{suppfig:swap}    
\end{figure}

Additionally, unlike n-type and p-type CMOS, there is only one type of SuperMag switch and it passes V$^+$ and V$^-$ with equal strength.  V$^+$ and V$^-$ can therefore be swapped without adverse consequences, enabling non-inverting gates (as illustrated in the AND gate of Supp. Fig. \ref{suppfig:swap}b).  This swapping is not possible in CMOS.  Input current direction can also be flipped by swapping the input port and ground of the switching wire, effectively inverting the logical value of a gate input (as illustrated in the OR and NOR gates of Supp. Fig. \ref{suppfig:swap}c,d).

Furthermore, by swapping the direction of one of the inputs and not the other, four more logic functions are available: $\bar{\tn{A}}\tn{B}$, $\tn{A}\bar{\tn{B}}$, $\bar{\tn{A}}+\tn{B}$, and $\tn{A}+\bar{\tn{B}}$. An immediate corollary of this is that unlike CMOS, SuperMag does not need dedicated inverters.  All inverting operations may be absorbed into upstream gate outputs or downstream gate inputs with zero overhead.  Finally, since a single switch layout can be reused for numerous logic functions, minimal standard cell design is needed for SuperMag.

\pagebreak

\begin{figure}
    \centering
    \includegraphics[width=\linewidth]{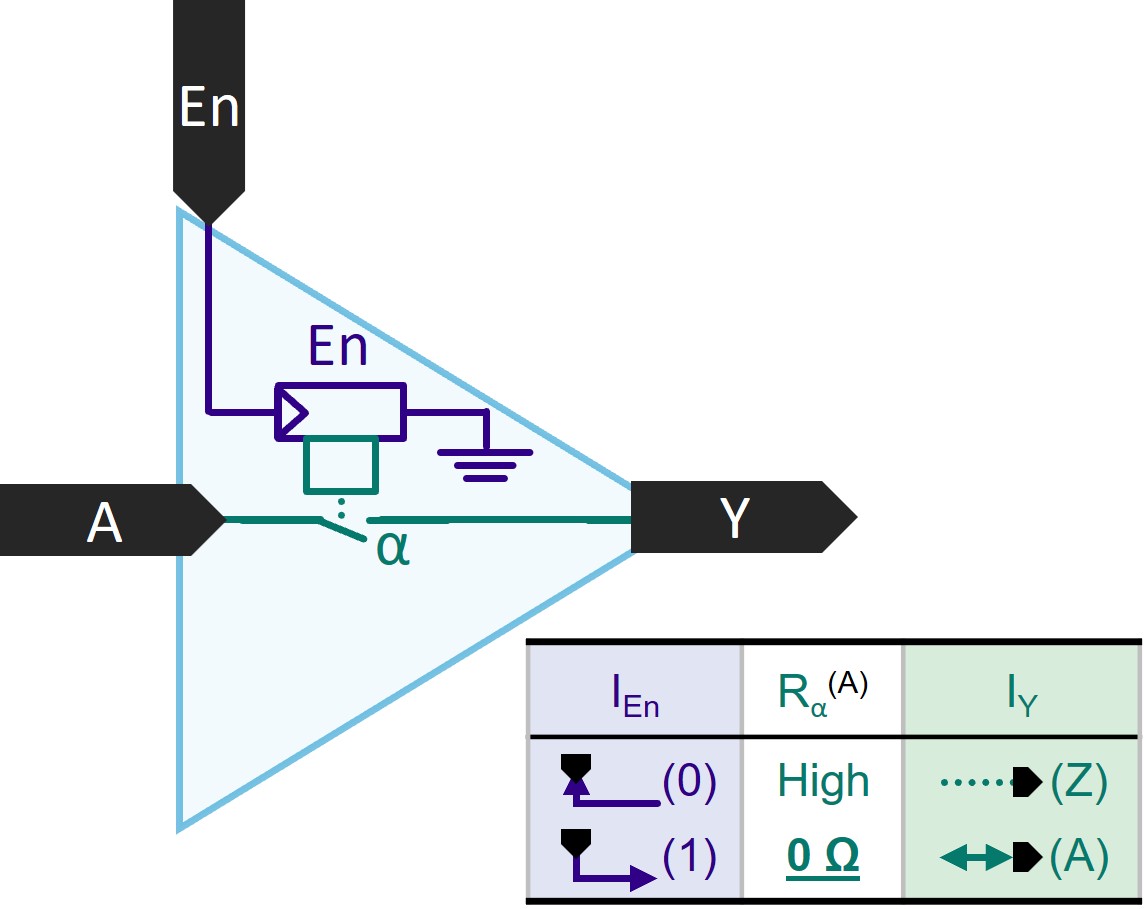}
    \caption{SuperMag tri-state buffer.  In contrast to CMOS transistors, SuperMag switches make excellent transmission gates with no signal loss because the ON-state has zero resistance.}
    \label{fig:tristate}
\end{figure}

\begin{figure}
    \centering
    \includegraphics[width=0.9\linewidth]{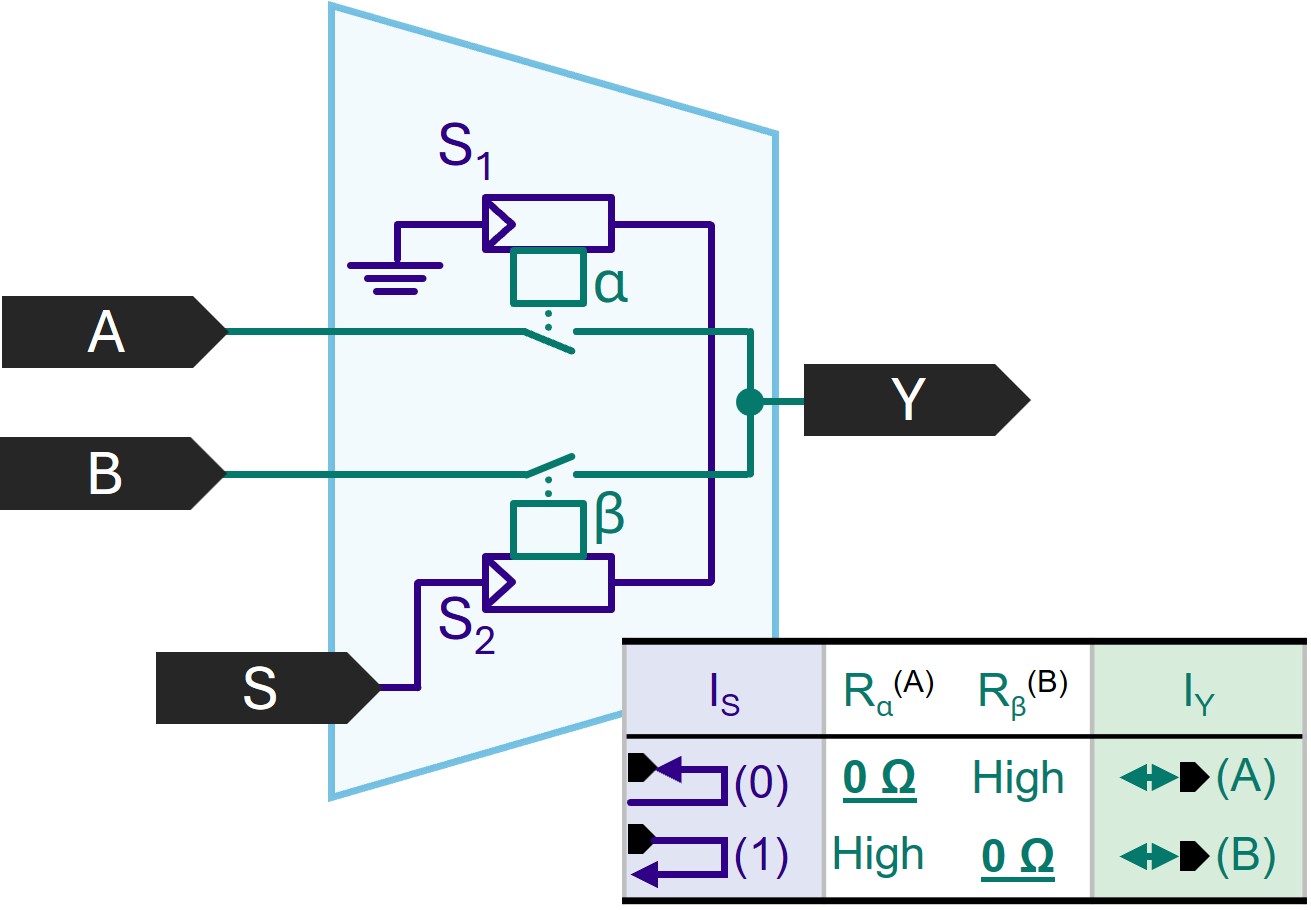}
    \caption{SuperMag MUX.  The MUX comprises two SuperMag switches used to select which input gets reflected at the output.}
    \label{fig:mux}
\end{figure}

\begin{figure}
    \centering
    \includegraphics[width=\linewidth]{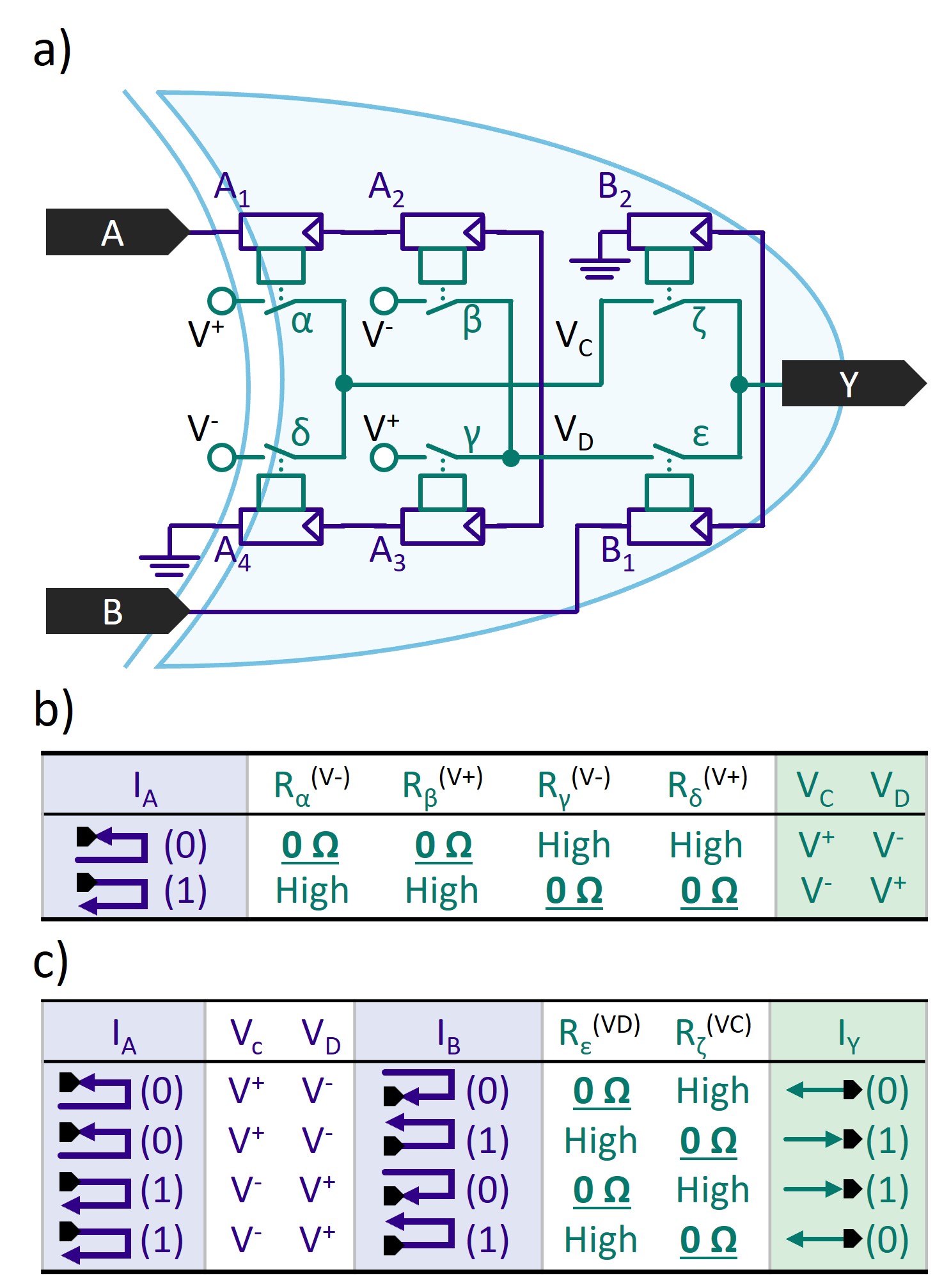}
    \caption{Complementary SuperMag XOR.  a) Gate schematic. b) Truth table for internal nodes V$_C$ and V$_D$.  c) XOR truth table.}
    \label{fig:xor}
\end{figure}

\section{\\SuperMag Transmission Gate Circuits}
\label{supnote:tgates}

SuperMag switches are perfect transmission gates: when turned on, there is a zero-resistance channel connecting both sides of the output wire.  This is in direct contrast to transistors, where the on-resistance of the source-drain path is non-negligible, necessitating frequent regeneration of signal strength, and strength depends on the driven logic value, necessitating two transistors for a good transmission gate.  These perfect SuperMag transmission gates significantly reduce gate overhead compared to CMOS and enable elegant solutions for arithmetic optimization.

\subsection*{Tri-State Buffer}

As SuperMag switches are perfect transmission gates, a single SuperMag switch can create a tri-state buffer, as illustrated in Supp. Fig. \ref{fig:tristate}.  When switch $\alpha$ is closed (following a logic `1' from input En), output Y is driven directly by input A.  In contrast, when switch $\alpha$ is open, output Y is electrically disconnected from input A, completing the tri-state functionality, which normally requires at least three transistors in CMOS. 

\subsection*{Multiplexer}
\label{sec:mux}

 A full-swing, full-drive multiplexer (MUX) can be constructed from two SuperMag switches acting as transmission gates as illustrated in Supp. Fig. \ref{fig:mux}.  The SuperMag switches are connected in a complementary manner, such that turning on one switch turns off the other, ensuring there are no zero-resistance paths between A and B.  When a logic `0' (`1') is presented to input S, switch S$_1$ turns on (off) and switch S$_2$ turns off (on), connecting output Y to input A (B).  Input S thus selects which input is passed to downstream logic, thereby performing the MUX function.  As a single SuperMag switch drives V$^+$ and V$^-$ with equal strength, only one switch is needed for each branch of the MUX.  In contrast, CMOS requires both a PMOS and NMOS to connect each input to the output with full strength, at the cost of significant overhead.

\subsection*{SuperMag XOR Gate}
\label{sec:xor}

As described in Supplementary Note \ref{supnote:cmos-map}, the conventional CMOS XOR circuit can be mapped one-to-one to the SuperMag logic family.  However, the unique physics of SuperMag switches enable the realization of XOR gates with only six SuperMag switches, as illustrated in Supp. Fig. \ref{fig:xor}a.  The four switches connected to input A generate intermediate voltages V$_C$ and V$_D$, with V$_C$ inverting input A and V$_D$ copying it.  The complementary switch pair connected to input B forms a MUX and selects whether output Y is connected to V$_C$ or V$_D$: if B is `0', Y connects to V$_D$ and outputs the value provided to A.  If B is `1', Y connects to V$_C$ and outputs the complement of the value provided to A.  The XOR functionality is thus achieved with only six switches, whereas the conventional complementary CMOS XOR gate requires at least ten transistors.  

\section{\\Additional Memory Circuits}
\label{supnote:secondary-memory}

Besides nvRAM, SuperMag's native non-volatility and perfect transmission gates enable elegant implementations of latches, D flip-flops, and configuration memory for programmable logic applications.

\subsection*{Transparent Latch}

A transparent latch is a memory circuit that can output a stored value Q provided via input D.  When the enable signal En is logic `1', the value that is input at D is transferred to Q.  When En is `0', the output Q is the value provided to D immediately before En was lowered.  Transparent latches are ubiquitous in conventional CMOS logic. 

\begin{figure}
    \centering
    \includegraphics[width=\linewidth]{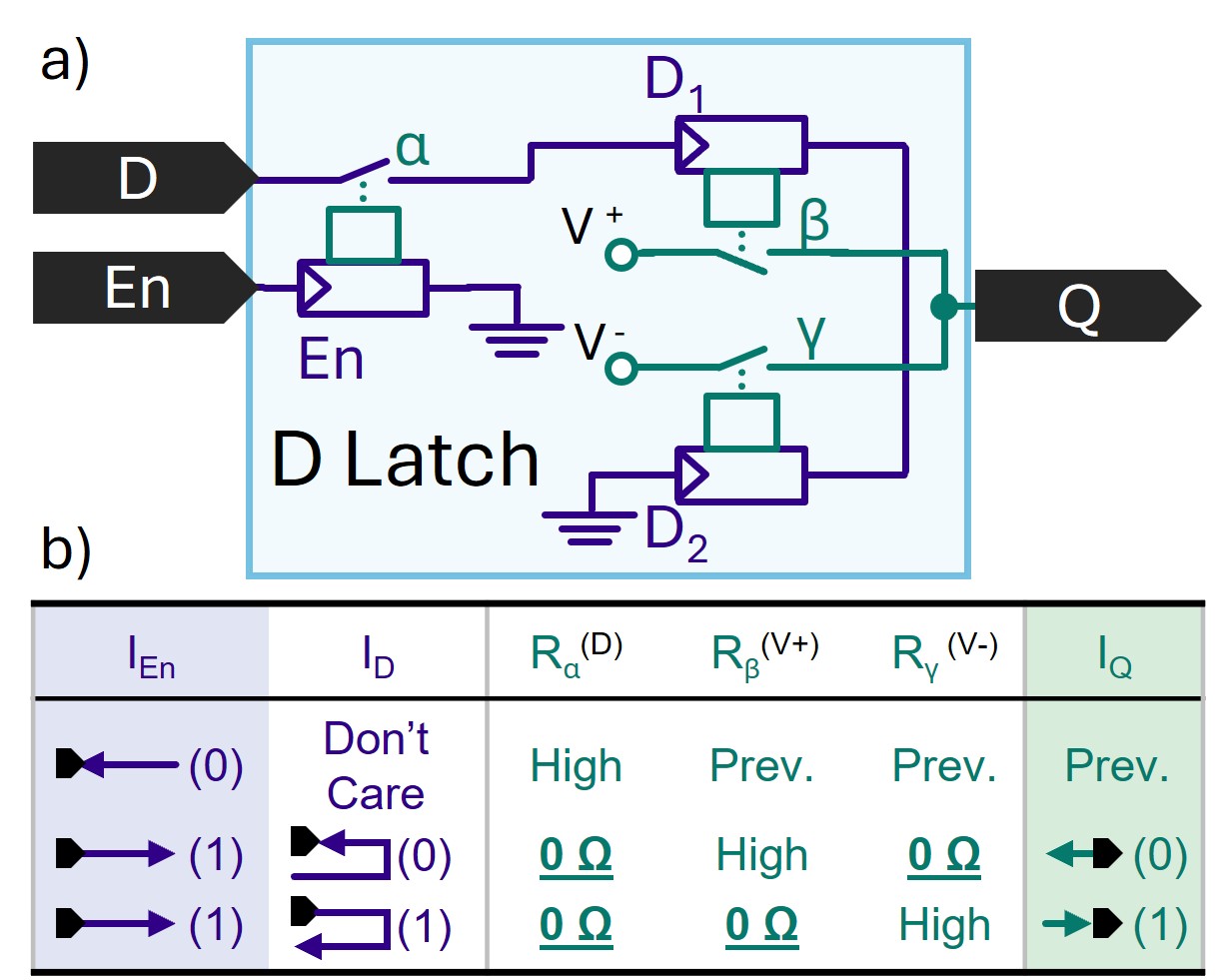}
    \caption{SuperMag transparent latch schematic and truth table. When enabled, the single transmission gate on the left allows input D to overwrite the state stored in the SuperMag buffer on the right, causing the output Q to reflect D.  Due to the non-volatility of SuperMag buffer, when the gate is disabled (En = `0'), the output retains the previous state.}
    \label{fig:dlatch}
\end{figure}

A three-switch SuperMag transparent latch can be constructed by combining the SuperMag transmission gate with the native non-volatility of SuperMag.  As illustrated in Fig. \ref{fig:dlatch}, the latch is formed from a transmission gate on the left and a SuperMag complementary buffer on the right.  When En is `1', switch $\alpha$ closes, allowing current from input D to pass to the input of the SuperMag buffer and overwrite its state.  Output Q therefore reflects input D when En is `1'.  When En is `0', switch $\alpha$ opens, creating a high resistance that impedes current from input D.  Any residual current flowing through the switching wire of the buffer from input D to ground is insufficient to switch the buffer, which retains its previous state due to its non-volatility.  The circuit thus provides the latch functionality. Robust CMOS designs generally require at least eight transistors to achieve the same functionality.

\begin{figure}
    \centering
    \includegraphics[width=0.8\linewidth]{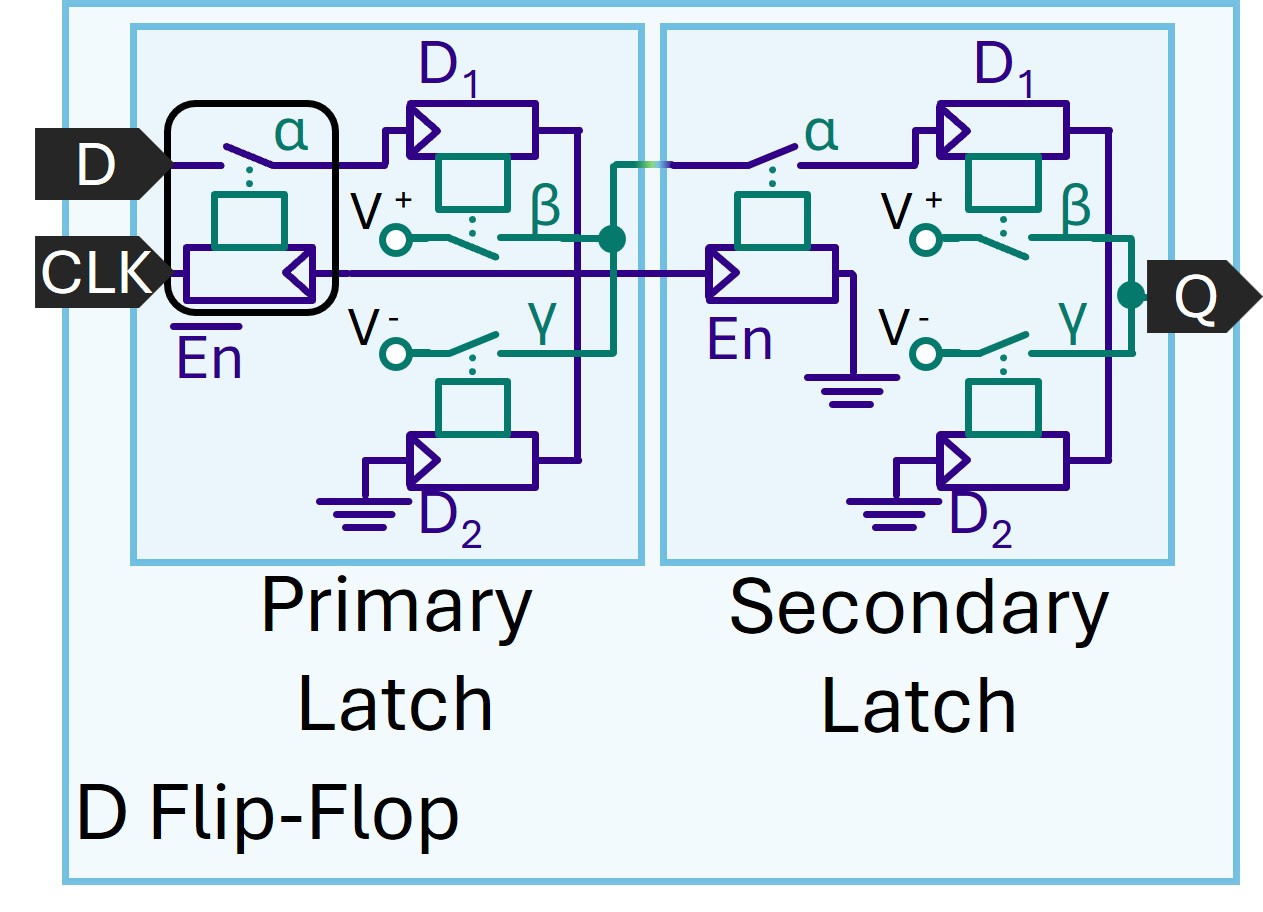}
    \caption{SuperMag D flip-flop constructed from two transparent latches.  The circled SuperMag switch has been reversed, activating when clk = `0'.}
    \label{fig:dff}
\end{figure}

\begin{figure}
    \centering
    \includegraphics[width=0.8\linewidth]{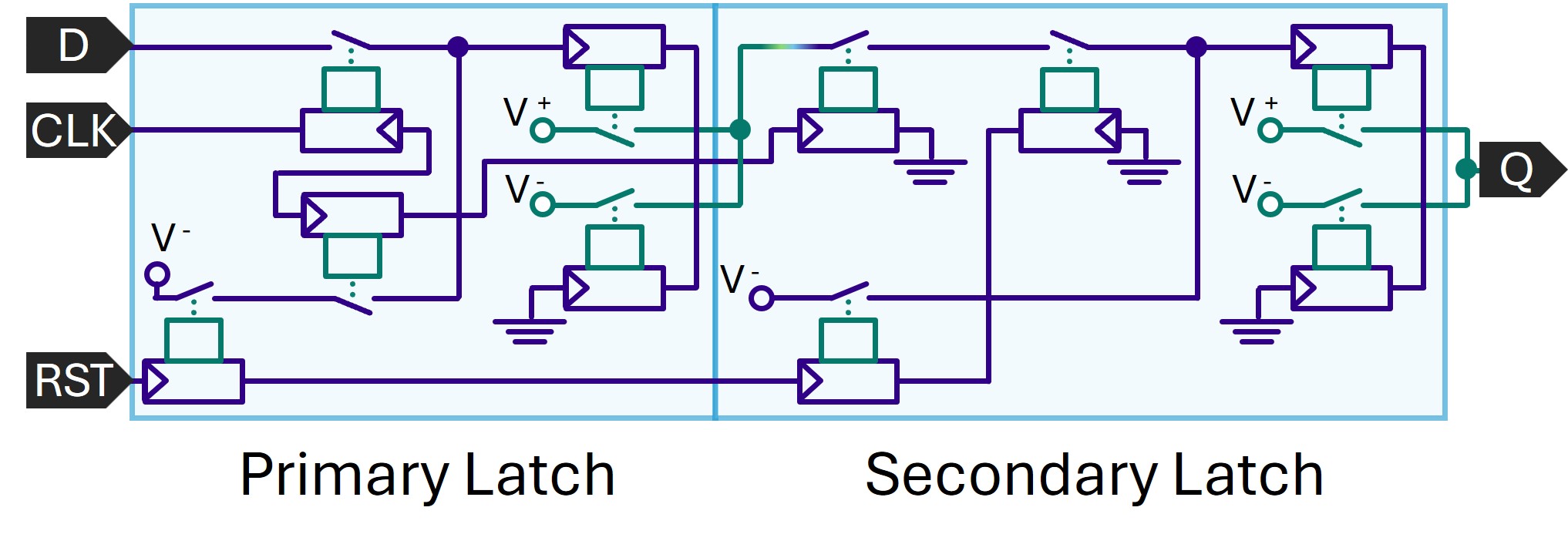}
    \caption{SuperMag D flip-flop with asynchronous reset.}
    \label{fig:dffr}
\end{figure}

\begin{figure}
    \centering
    \includegraphics[width=0.8\linewidth]{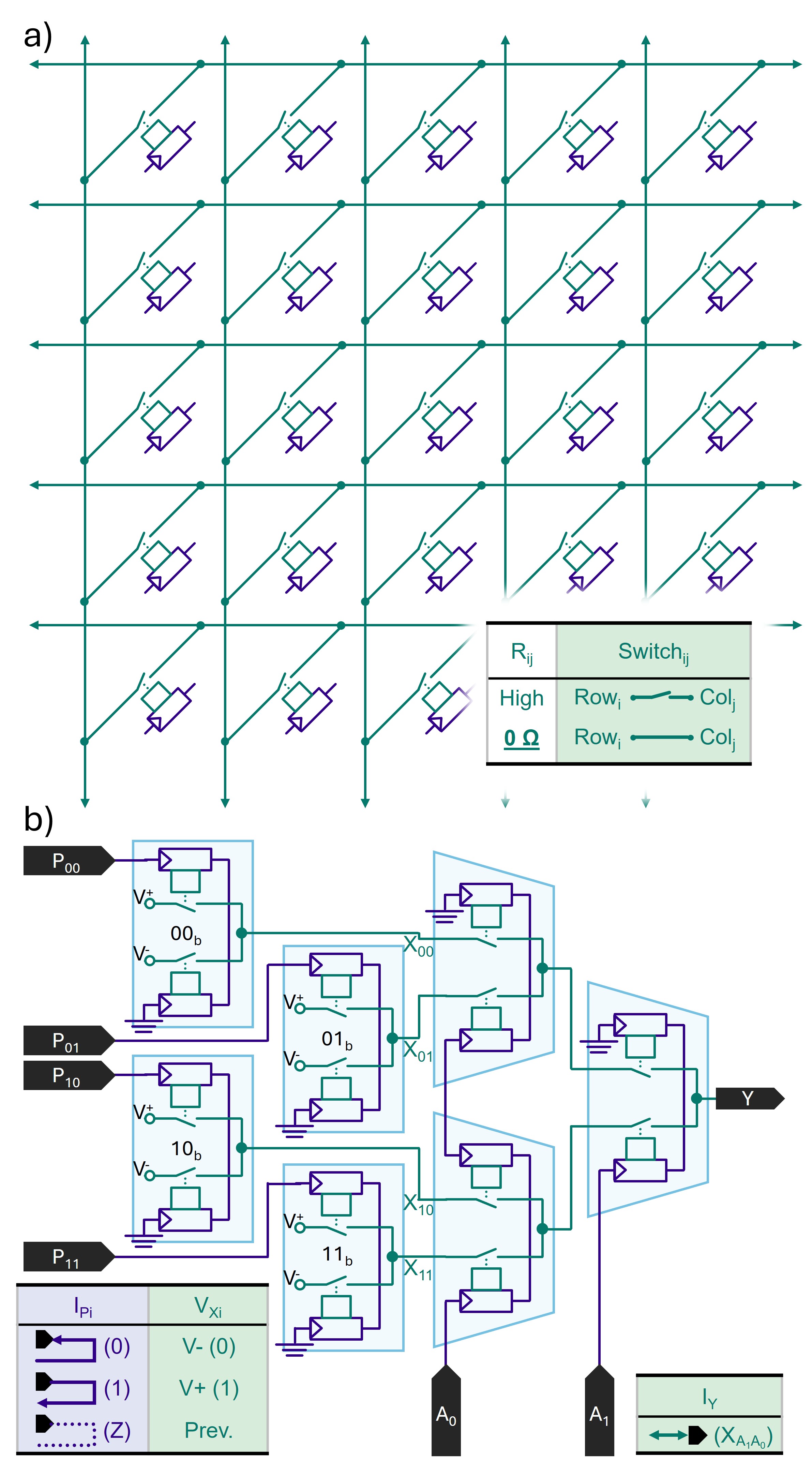}
    \caption{Re-programmable logic. a) Routing in a programmable array can be configured using non-volatile transmission gates. b) Logic look-up tables (LUTs) can be built from SuperMag MUXs and non-volatile latches. }
    \label{fig:fpga}
\end{figure}

\subsection*{D Flip-Flop}
Two daisy-chained transparent latches with complementary En signals comprise a D flip-flop, the workhorse memory element central to sequential logic.  The D flip-flop latches data only on the \textit{rising edge} of the enable or clock signal and is never transparent, effectively isolating Q from D at all times other than the rising edge of the clock.

As in CMOS, two SuperMag transparent latches can be used to create a SuperMag D flip-flop as illustrated in Fig. \ref{fig:dff}.  Though conventional latch-based D flip-flops require an extra inverter to invert one of the two enable signals, the SuperMag D flip-flop does not need these additional devices as a result of the bidirectional nature of the SuperMag system.   The master latch intrinsically provides this inverting functionality by flipping the En SuperMag switch, effectively reversing the CLK input direction of that switch as described in Section \ref{sec:nand}a.  Additionally a D flip-flop with asynchronous reset can be constructed as illustrated in Supp. Fig. \ref{fig:dffr}.  This ten-SuperMag-switch circuit normally requires 20 transistors in CMOS.

\subsection*{Reconfigurable Hardware}

Combining SuperMag non-volatility and transmission gate functionality enables very-low-footprint reconfigurable hardware.  This is in direct contrast to modern CMOS FPGAs, where programming bitstreams are stored externally to the reconfigurable logic, necessitating significant latency upon startup.  Fig. \ref{fig:fpga}a illustrates reconfigurable routing, utilizing a crossbar structure of SuperMag transmission gates that enable or disable connections between individual rows and columns.  Fig. \ref{fig:fpga}b depicts a two-input SuperMag lookup table.  Programming bits are stored in the non-volatile buffers on the left-hand side of the figure and selected by logic via the multiplexers on the right.

\pagebreak

\section{\\Functionality and Energy Efficiency}
\label{supnote:sm_calcs}

SuperMag has a number of material system requirements that must be satisfied for SuperMag to be functional and energy efficient.  The following subsections describe these constraints.

\subsection*{Functionality}
\label{sec:functionality}

There are two main functional constraints on potential SuperMag material systems: 

\begin{itemize}
    \item current driven through the SC must be sufficient to switch downstream SuperMag switches and
    \item resistance through the SOT layer of downstream logic should be much less than the normal off-resistance of the disabled SuperMag switch in a complementary pair.
\end{itemize}

\noindent Restated formally,

\begin{subequations}
    \begin{align}
        I_c > I_{sot} \label{eq:const_i}\\
        R_{sc} >> R_{sot}. \label{eq:const_r}
    \end{align}
\end{subequations}

\begin{figure}[h]
    \centering
    \includegraphics[width=0.6\linewidth]{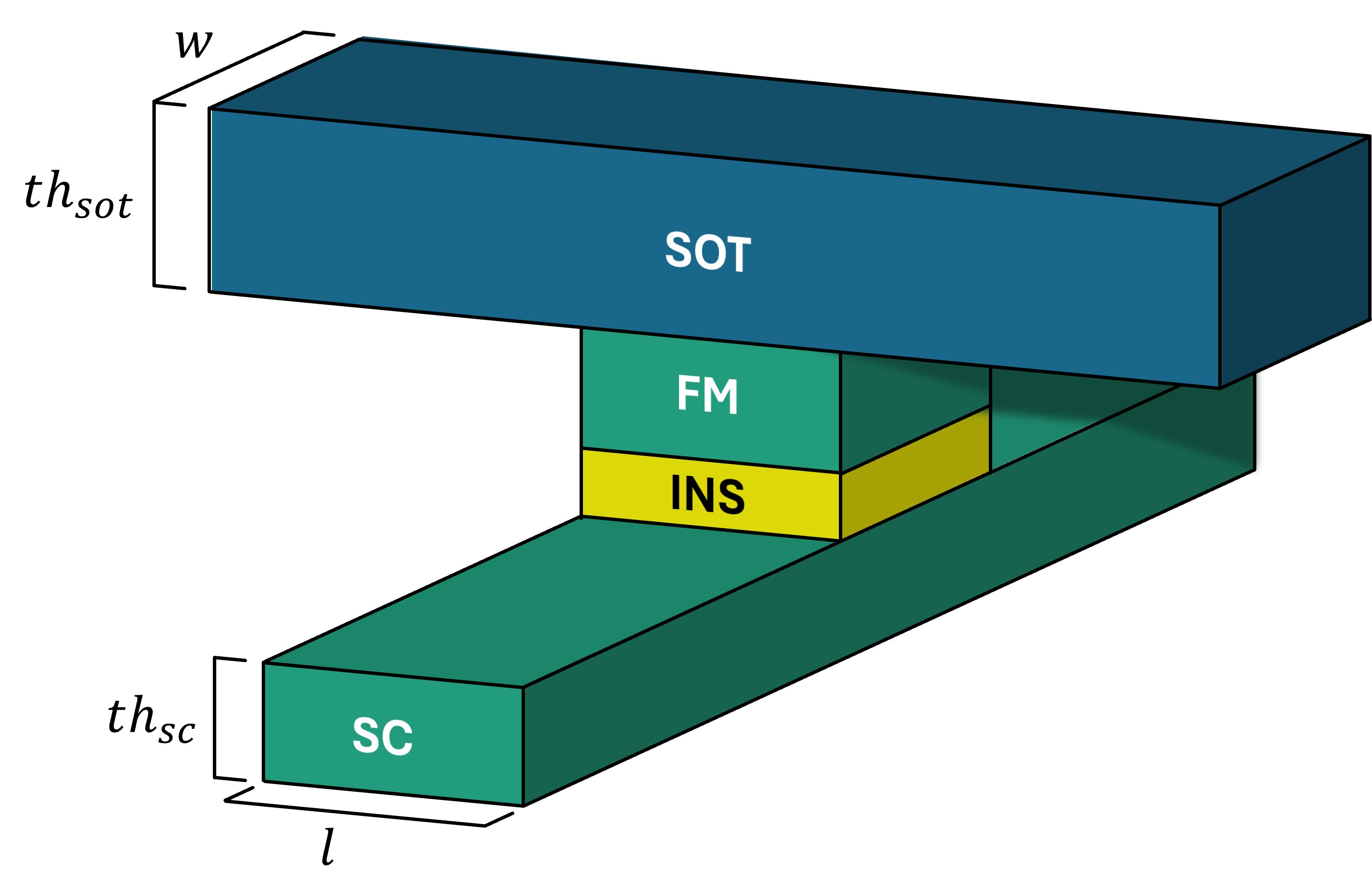}
    \caption{Critical dimensions of the SuperMag switch.}
    \label{fig:dim-perp}
\end{figure}

\noindent A 3-dimensional illustration of the SuperMag gate is depicted in Supp. Fig. \ref{fig:dim-perp} along with marked dimensions of the switch, $w$, $l$, $th_{sot}$, and $th_{sc}$.  In addition, the SC has critical current density $J_c$ and normal resistivity of $\rho_{sc}$, and the SOT layer has needed switching current density of $J_{sot}$ and resistivity of $\rho_{sot}$.  For the current constraint (\ref{eq:const_i}),

\begin{equation}
    I_c = J_c \times l \times th_{sc}
\end{equation}

\noindent and

\begin{equation}
    I_{sot} = J_{sot} \times w \times th_{sot}.
\end{equation}

\noindent substituting these expressions into (\ref{eq:const_i}) gives,

\begin{equation}
    \frac{J_c}{J_{sot}} > \frac{w \times th_{sot}}{l \times th_{sc}} \label{eq:const_j}.
\end{equation}

\noindent Similarly, as,

\begin{equation}
    R_{sc} = \rho_{sc} \times \frac{w}{l \times th_{sc}}
\end{equation}

\noindent and

\begin{equation}
    R_{sot} = \rho_{sot} \times \frac{l}{w \times th_{sot}},
\end{equation}

\noindent the resistance constraint (\ref{eq:const_r}) may be rewritten as,

\begin{equation}
    \frac{w^2 \times th_{sot}}{l^2 \times th_{sc}} >> \frac{\rho_{sot}}{\rho_{sc}}. \label{eq:const_rho}
\end{equation}

\noindent Substituting (\ref{eq:const_j}) into (\ref{eq:const_rho}) gives

\begin{equation}
    \frac{w}{l} \times \frac{J_c}{J_{sot}} >> \frac{\rho_{sot}}{\rho_{sc}} \Rightarrow \frac{w}{l} >> k_{SuperMag},
    \label{eq:wl_rat}
\end{equation}

\noindent where $k_{SuperMag} = (J_{sot}\rho_{sot}) / (J_c\rho_{sc})$.  Substituting back into (\ref{eq:const_j}) gives,

\begin{equation}
    \frac{J_c}{J_{sot}} >> k_{SuperMag} \times \frac{th_{sot}}{th_{sc}} \Rightarrow \frac{th_{sc}}{th_{sot}} >> k_{SuperMag} \times \frac{J_{sot}}{J_c}. \label{eq:th_rat}
\end{equation}

\noindent Assuming that the thickness ratio $th_{sc} / th_{sot}$ is no greater than five and the current density ratio $J_c / J_{sot}$ is no greater than two gives: 

\begin{equation}
    k_{SuperMag} = \frac{J_{sot}\rho_{sot}}{J_c\rho_{sc}} << 10.
\end{equation}

\noindent The SuperMag constraints therefore favor switches that have a large width to length ratio where the thickness of the SC is large compared to the thickness of the SOT layer.  These minimum target ratios are governed by material properties (current density and resistivity of the SC and SOT layers), and materials should be chosen to minimize $k_{SuperMag}$.  Particularly, $J_{sot}$ and $\rho_{sot}$ should be minimized, and $J_c$ and $\rho_{sc}$ should be maximized in both cases.

Supplementary Tables \ref{tab:sc} through \ref{tab:func} summarize various material systems in the literature and highlight potential candidate material pairs for SuperMag.  Material systems with Nb or NbN have much lower $k_{SuperMag}$ \cite{SOT}, and a SOT material with low $J_{sot}$ such as Bi$_3$Sb$_2$ \cite{sot-bisb} or PtHf \cite{SOT} are also favored.

\setul{1pt}{.4pt}

\begin{table}[t]
    \centering
\caption{\textbf{{Potential SuperMag Superconducters}}}
\begin{tabular}{|l|c|cccc|} \hline
    \textbf{Material System}   & \textbf{Study/Note}  & \textbf{$\mathbf{J}_\mathbf{c}$ (A/m$^2$)} & \textbf{$\rho_\mathbf{sc}$ ($\Omega \mathbf{m}$)} & \textbf{$\mathbf{J}_\mathbf{c}\rho_\mathbf{sc}$ (V/m)} & \textbf{$\mathbf{J}_\mathbf{c}^\mathbf{2}\rho_\mathbf{sc}$ (W/m$^\mathbf{3}$)} \\
     \hline 
    Al                         &  this work           &                $1.1\times10^{10}$          &            $1\times10^{-7}$                       & $1.1\times10^{3}$                                      & $1.2\times10^{13}$          \\
    Pb                         &  \cite{sc_pb}        &                $2\times10^{10}$            &            $1\times10^{-7}$                       & $2\times10^{3}$                                        & $4\times10^{13}$              \\
    Nb                         &  \cite{sc_nb}        &                $4\times10^{10}$            &            $6\times10^{-7}$                       & $2.4\times10^{4}$                                      & $9.6\times10^{14}$              \\
    NbN                        &  \cite{SOT}          &                $2.5\times10^{10}$          &            $3\times10^{-6}$                       & $7.5\times10^{4}$                                      & $1.9\times10^{15}$              \\
     \hline
     
\end{tabular}
\label{tab:sc}   

\end{table}

\begin{table}[t]
    \centering
        \caption{\textbf{{Potential SuperMag SOT/FM Stacks}}}
        \begin{tabular}{|l|c|cccc|} \hline
             \textbf{Material System}  & \textbf{Study}    & \textbf{$\mathbf{J}_\mathbf{sot}$ (A/m$^\mathbf{2}$)} & \textbf{$\rho_\mathbf{sot}$ ($\Omega \mathbf{m}$)} & \textbf{$\mathbf{J}_\mathbf{sot}\rho_\mathbf{sot}$ (V/m)} & \textbf{$\mathbf{J}_\mathbf{sot}^\mathbf{2}\rho_\mathbf{sot}$ (W/m$^\mathbf{3}$)}  \\
             \hline 
            Pt/CoFeB                   &  \cite{sot4}      &       $5\times10^{11}$                                &       $3.4\times10^{-7}$                           & $1.7\times10^{5}$                                         & $8.5\times10^{16}$ \\
            {[Pt/Co]}x/CoFeB           &  \cite{sot-ptco}  &       $3.7\times10^{11}$                              &       $3\times10^{-7}$                             & $1.1\times10^{5}$                                         & $4.1\times10^{16}$ \\
            Bi$_3$Sb$_2$/CoPt          &  \cite{sot-bisb}  &       $1.5\times10^{10}$                              &       $6.7\times10^{-6}$                           & $1.0\times10^{5}$                                         & $1.5\times10^{15}$ \\
            PtHf/CoFeB                 &  \cite{SOT}       &       $7.5\times10^{10}$                              &       $8\times10^{-7}$                             & $6.0\times10^{4}$                                         & $4.5\times10^{15}$ \\
            Ta/CoFeB                   &  \cite{sot2}      &       $4\times10^{12}$                                &       $1.3\times10^{-7}$$^*$                       & $5.2\times10^{5}$                                         & $2.1\times10^{18}$ \\
             \hline
             
        \end{tabular}
        \label{tab:sot}
        \begin{tablenotes}
            \item $^*$Unspecified in work. Using common-knowlege estimate.
        \end{tablenotes}
\end{table}

\begin{table}[b]
    \centering
\caption{\textbf{{Potential SuperMag Material Systems}}}
\begin{tabular}{|l|c|cc|} \hline
       &                 & \textbf{$\mathbf{k}_\mathbf{SuperMag}$} & \textbf{$\mathbf{k}_\mathbf{SuperMag} \mathbf{J}_\mathbf{sot}/\mathbf{J}_\mathbf{c}$}   \\
    \textbf{Material System}     & \textbf{Study}  & \textbf{=min(w/l)}     & \textbf{=min($\mathbf{th}_\mathbf{sc}/\mathbf{th}_\mathbf{sot}$)}   \\
     \hline 
    Pt/CoFeB/MgO/Al                &  \cite{sot4}, this work        & 155          &  7080            \\ 
    Bi$_3$Sb$_2$/CoPt/(Ins)/Al             &  \cite{sot-bisb}, this work    & 91           &  1250            \\
    PtHf/CoFeB/MgO/Al              &  \cite{SOT}, this work         & 55           &  375             \\
    Bi$_3$Sb$_2$/CoPt/(Ins)/Pb             &  \cite{sot-bisb}, \cite{sc_pb} & 50           &  38              \\
    PtHf/CoFeB/(Ins)/Pb            &  \cite{SOT}, \cite{sc_pb}      & 30           &  113             \\
    Bi$_3$Sb$_2$/CoPt/(Ins)/Nb             &  \cite{sot-bisb}, \cite{sc_nb} & 4.2          &  1.6             \\
    PtHf/CoFeB/(Ins)/Nb            &  \cite{SOT}, \cite{sc_nb}      & 2.5          &  4.7             \\
    \textbf{BiSb/CoPt/(Ins)/NbN }  &  \cite{sot-bisb}, \cite{SOT}   & 1.3          &  \textbf{0.79 }  \\
    \textbf{PtHf/CoFeB/(Ins)/NbN } &  \cite{SOT}                    & \textbf{0.8} &  2.4             \\
     \hline
     
\end{tabular}
\label{tab:func}

\end{table}

\subsection*{Energy Efficiency}

\begin{figure}
    \centering
    \includegraphics[width=0.7\linewidth]{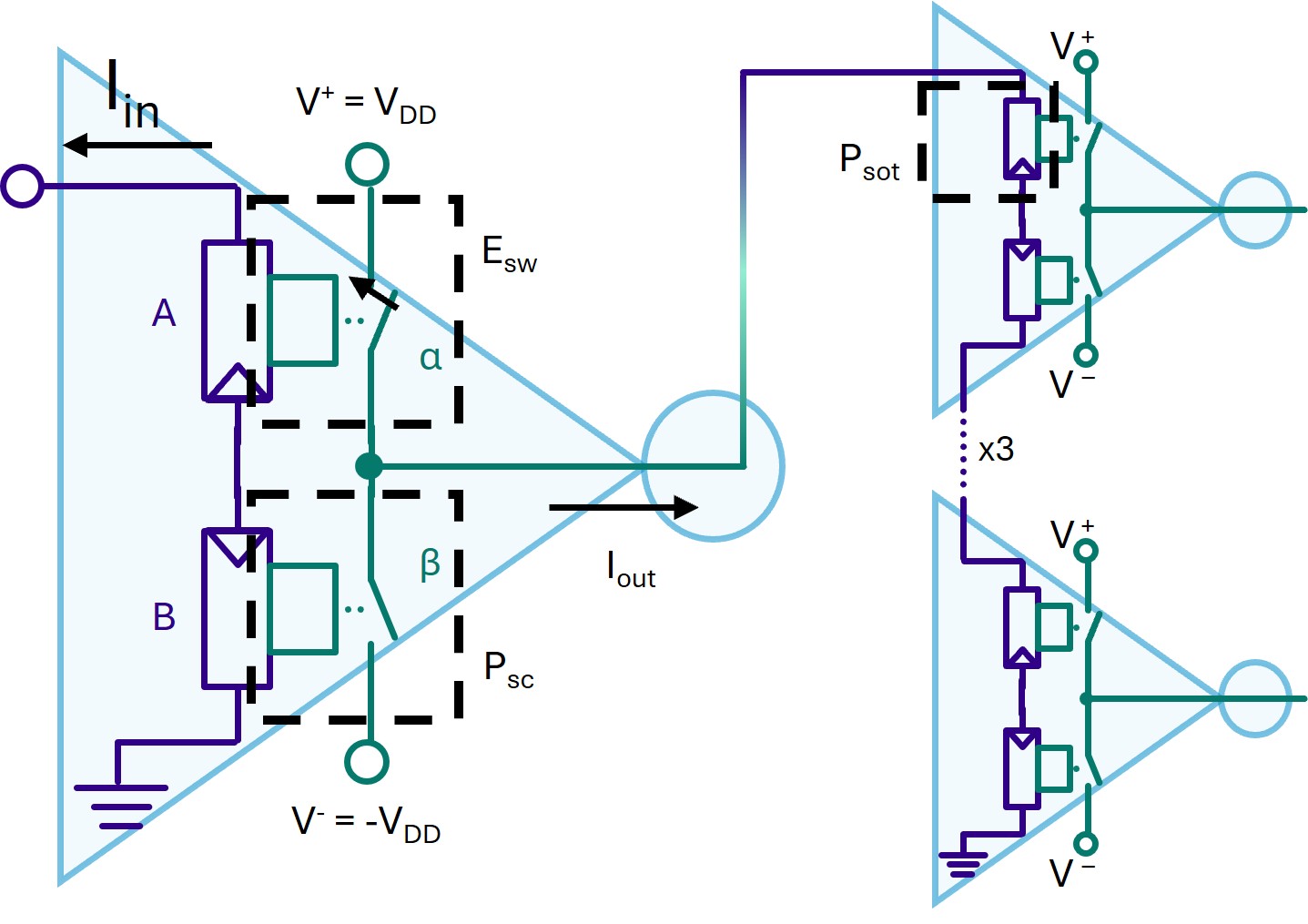}
    \caption{Schematic of the circuit used to estimate energy and power dissipation.}
    \label{fig:schem}
\end{figure}

Supp. Fig. \ref{fig:schem} gives the schematic used to approximate power and energy consumption of SuperMag.  V$_\tn{DD} =$ V$^\tn{+}= -$V$^\tn{-}$ is calculated as the minimum voltage needed to switch five downstream SuperMag inverters from the output of a SuperMag inverter.  The number five here is chosen arbitrarily to reasonably limit V$_\tn{DD}$ while minimally impacting logic design.  As fan-out is accomplished by stringing the switching wires of SuperMag switches in series, V$_\tn{DD}$ -- divided equally across the ten SuperMag switches of the five inverters -- will need to supply enough current to switch the SuperMag switches.  This is expressed as,

\begin{equation}
    V_{DD} = 10 \times I_{sot} \times R_{sot} = 10 \times J_{sot} \times \rho_{sot} \times l,
\end{equation}

\noindent which surprisingly does not depend on the switch width or thickness.  Assuming $V^+$ and $V^-$ are constantly presented to the circuit and not clocked (see \cite{FriedmanCMAT1}), there are three sources of energy dissipation in SuperMag, the switching energy of a SuperMag switch (\tn{E}$_\tn{sw, unit}$), static power leaked through downstream SOT layers after switching is completed (\tn{P}$_\tn{sot}$), and static power through the off SuperMag switch in a complementary pair (\tn{P}$_\tn{sc}$).  \tn{E}$_\tn{sw, unit}$ is calculated as,

\begin{equation}
    E_{sw, unit} = I_{sot}^2 \times R_{sot} \times t_{sw} = J_{sot}^2 \times \rho_{sot} \times Vol_{sot} \times t_{sw},
\end{equation}

\noindent where $t_{sw}$ is the average switching time of the SuperMag switch and $Vol_{sot}$ is the volume of the SOT portion of the switch: $Vol_{sot} = w \times l \times th_{sot}$.  Assuming the parameters of Bi$_3$Sb$_2$/CoPt/(Ins)/NbN in Supplementary Table \ref{tab:func}, a $Vol_{sot}$ of 12,500 nm$^3$, and a $t_{sw}$ of 2 ns, $E_{sw} \approx 50$ aJ.  $P_{sot, fo10}$ is calculated as,

\begin{equation}
    P_{sot, fo10} = V_{DD} \times I_{sot} = 10 \times J_{sot}^2 \times \rho_{sot} \times Vol_{sot},
\end{equation}

\noindent which is again proportional to the switch volume.  Finally, $P_{sc}$ is calculated as,

\begin{equation}
    P_{sc, inv} = \frac{(V^+ - V^-)^2}{R_{sc}} = \frac{(2 \times (10 \times J_{sot} \times \rho_{sot} \times l))^2}{\rho_{sc} \times w / (l \times th_{sc})} = 400 \times J_{sot}^2 \frac{\rho_{sot}^2}{\rho_{sc}} \times \frac{l^3 \times th_{sc}}{w}. \label{eq:psc1}
\end{equation}

\noindent Combining (\ref{eq:psc1}) with (\ref{eq:th_rat}) above and assuming inequality (\ref{eq:const_i}) is near equality gives:

\begin{multline}
    P_{sc, inv} = 400 \times J_{sot} \times J_c \times \rho_{sot} \times k_{SuperMag} \times Vol_{sot} \times \frac{th_{sc}}{th_{sot}} \times \frac{l^2}{w^2} \Rightarrow \\ P_{sc} << 400 \times J_{sot}^2 \times \rho_{sot} \times Vol_{sot}.
\end{multline}

\noindent Should a clocked voltage scheme be used as in \cite{FriedmanCMAT1}, $P_{sot}$ can be ignored and total energy consumption per clock cycle is approximated as:

\begin{equation}
    E_{total} = E_{sw, total} + P_{sc, total} \times t_{sw}.
\end{equation}

\noindent Whether using a level or clock-based scheme, SuperMag energy dissipation therefore scales directly with $J_{sot}^2\rho_{sot}Vol_{sot}$ and surprisingly does not depend on the SC parameters -- though the SC parameters are given great consideration earlier in this note.  $J_{sot}$ and $\rho_{sot}$ should therefore be minimized, agreeing with the functional constraints presented earlier in this note.  Additionally, as the energy dissipation is proportional to SuperMag switch volume, scaling down the whole device -- which will not affect the constraints of the previous subsection -- will reduce power consumption cubically, though manufacturing and thermal constraints will limit spatial scaling.\vfill

\section{\\System-Level Comparisons}
\label{supnote:comp}

As depicted in Fig. \ref{fig:comp}b-e of the main text, system-level comparisons were performed between CMOS, RSFQ, and SuperMag on a combinational full adder, sequential 32-bit counter, and full RISC-V CPU plus 2 KB of RAM.  All three designs (minus the RAM array) were synthesized in Cadence genus with technology-specific implementation details described in the following subsections.  The RISC-V processor is an Ibex RISC-V Core \cite{ibex} with default options.

\subsection*{CMOS}

The three designs were synthesized in Cadence Genus using the SkyWater Open Source Process Design Kit (PDK) sky130\_fd\_sc\_lp libraries \cite{sky130}.  Target clock frequencies were reduced until there were no timing violations, then area and power metrics were reported.

\textit{Area --} Area and standard cell count were reported by genus.  Device count for each standard cell was extracted from simulation spice files in the SkyWater 130 library.

\textit{Power --} Static power was directly pulled from the power report.  Switching power reported in the main text is the sum of the internal and switching powers reported by genus.  As SkyWater 130 is an older technology, switching power dominates.

\textit{Delay -- } Delay was computed as the reciprocal of the target clock period.  For the full adder, input and output registers were included so genus would properly compute this metric, then, the delay of the maximum path was used instead of the clock period for the delay metric.

\textit{RAM -- } CMOS RAM area and static power numbers come from the SKY130 Standard SRAM Configurations project \cite{sky130ram}, with area extracted from the provided GDSII file and static power from the Liberty file. 

\subsection*{RSFQ}

The ColdFlux RSFQ Logic Cell Library \cite{rsfqlib} was used to estimate representative area, power, and delay metrics for the RSFQ logic family.  Bias currents for the standard cells in \cite{rsfqlib} are supplied by applying a constant 2.6 mV across bias resistors. In order to get genus to synthesize in RSFQ, a custom Liberty file was created containing all of the RSFQ logic cells. As RSFQ is clocked at the cell level instead of the conventional RTL level, two clock domains are used, the 50 GHz \cite{rsfqlib} cell-level clock and the slower register-level clock.  The fast clock is communicated to genus by setting the delay of each gate in the Liberty file to 1 (\textit i.e. one fast clock cycle).  Additionally, splitter cells were not included in the Liberty file, and instead, inter-gate fan-out was limited to two (2) assuming a maximum of one splitter could fit between gates before violating the fast clock timing.  Once designs were synthesized, the synthesized netlist was used to determine the number of splitter cells used, the number of delay cells needed to synchronize the logical fan-in cones, and the number of splitters needed to clock all of the cells.  Finally area, power, and delay were estimated by combining this information and data from the cell and timing reports.

\textit{Area -- } The cell report gives the count of each standard cell in the synthesized design, and analysis of the netlist gives the number of splitter and delay DFF cells needed.  The number of Josephson junctions (JJs) for each standard cell was extracted from the simulation spice files in \cite{rsfqlib}, and standard cell area was extracted from the provided GDSII files \cite{rsfqlib}.  

\textit{Power -- } Static power for each of the standard cells was obtained by summing up the total bias current of the cell and multiplying by the 2.6 mV across the bias resistors.  Average switching power for each of the standard cells was extracted by simulating the JoSIM \cite{josim} test benches provided in \cite{rsfqlib} to obtain average switching energy per clock cycle and multiplying by the fast clock frequency of 50 GHz.  

\textit{Delay -- } The slow clock period is the maximum number of fast-clock stages in a register path divided by the fast clock frequency of 50 GHz \cite{rsfqlib}.

\textit{RAM -- } To estimate the area and power of a 2 KB JJ-based RAM, an RSFQ address decoder was synthesized.  In addition, bit-cell area was estimated for an optimized vortex transitional (VT) cell to be 168 $\mu$m$^2$ \cite{rsfqbitcell}, and the area of the 2 KB array (16,384 cells) is directly computed.  Array static power is 0 as data is stored as superconducting current in a loop, and array switching power is negligible as only one word of the array switches for any access.  Finally, area of the latching drivers and latching sense circuits, used to convert SFQ pulses to DC current and vice versa, is negligible, and static power of the latching drivers is computed as follows: latching drivers supply 1 mA to the activated word line and every bit line during an access, requiring that a periodic 1 mA signal (with assumed 75\% duty cycle) must be supplied to \textit{every} latching driver, incurring a static power cost.  It is assumed that this bias current is supplied by pulsing a 2.6 mV across a resistor as in \cite{rsfqlib}, and a P$_{stat, driver}$ = 1 mA $\times$ 0.75 $\times$ 2.6 mV = 1.95 $\mu$ W is therefore incurred for each row and column in the array.  

\subsection*{SuperMag}

As with the RSFQ library, a custom liberty file was used to have genus synthesize with SuperMag gates.  As described in Section \ref{sec:logic-and-memory}, inverting inputs or outputs of combinational gates can be done without changing the cell structure, so all combinations of gates with inverted inputs and outputs were made available to genus.  Additionally, gate fan-out was limited using the output\_pin.max\_capacitance and input\_pin.capacitance liberty properties such that each gate is limited to driving at most 10 SuperMag switches in downstream logic.  The area of each cell was configured as the number of SuperMag switches (unitless) in that cell, so that the area reported by genus is the number of SuperMag switches in the design.  Worst-case delay was assumed to be the same for every cell.  Normalized values of $P_{sc}$ were included in the Liberty file as leakage, and normalized values of $E_{sw}$ for each gate input were included as internal power.  In addition to the full adder, counter, and processor, a 2 KB decoder was also synthesized to estimate RAM numbers.  After synthesis, the cell, timing, and power reports are post-processed to obtain area, power, and delay metrics.

Principal material and design rule parameters are specified in Supp. Table \ref{tab:sm_params}.  J$_\tn{c}$ and $\rho_\tn{sc}$ come from the NbN numbers of Supplementary Table \ref{tab:sc} and J$_\tn{sot}$ and $\rho_\tn{sot}$ from the Bi$_3$Sb$_2$/CoPt numbers of Supplementary Table \ref{tab:sot}.  Switching time (t$_\tn{sw}$) and minimum spacial dimensions (l$_\tn{min}$ and w$_\tn{min}$ were gleaned from \cite{sot_switch_time}.  k$_\tn{wl}$ is an additional factor included in (\ref{eq:wl_rat}) to ensure the strong inequality is not violated and k$_\tn{th}$ is similarly used to ensure inequality (\ref{eq:th_rat}) holds.  Safety factor sf$_\tn{area}$ is used to calculate the average footprint of a SuperMag switch with A $=$ w $\times$ l $\times$ sf$_\tn{area}$.  The fanout parameter specifies a maximum fanout of ten SuperMag switches can be driven from an upstream switch, enabling the optimization of V$_\tn{dd}$.  Parameters shaded in gray are derived from the other parameters according to the analysis of Supplementary Note \ref{supnote:sm_calcs}.

With the expectation of future maturation and optimization of materials, comparisons for SuperMag with more ideal parameters are depicted in Fig. \ref{fig:comp}b-e of the main text.  To compute these, t$_\tn{sw}$, J$_\tn{c}$, $\rho_\tn{sc}$, J$_\tn{sot}$, and $\rho_\tn{sot}$ were scaled by factor k$_\tn{opt}$ with J$_\tn{c}$, $\rho_\tn{sc}$ being increased and t$_\tn{sw}$, J$_\tn{sot}$, and $\rho_\tn{sot}$ decreased by the same factor.  In Fig. \ref{fig:comp}b-e k$_\tn{opt} =$ 5, and a k$_\tn{opt}$ of 1.32 is needed for SuperMag to match the processor PDP of CMOS.     

\setul{1pt}{.4pt}
\definecolor{gray1}{HTML}{E5E5E5}
\newcommand{\Gy}{\cellcolor{gray1}}

\begin{table}[t]
    \centering
\caption{\textbf{{SuperMag Material Parameters}}}
\begin{tabular}{|l|l||l|l||l|l|} \hline
    \textbf{Param}    & \textbf{Value}                          &    \textbf{Param}        & \textbf{Value}       & \textbf{Param}      & \textbf{Value}     \\ 
    \hline 
    t$_\tn{sw}$       &  0.19 ns                                &    th$_\tn{sc, min}$     &  5 nm                & \Gy th$_\tn{sc}$    & \Gy 67 nm          \\                                 
    J$_\tn{c}$        &  2.5$\times$10$^\tn{10}$ Am$^\tn{-2}$   &    f$_\tn{wl}$           &  10                  & \Gy V$_\tn{dd}$     & \Gy 30 mV          \\                                                   
    $\rho_\tn{sc}$    &  3.6$\times$10$^\tn{-6}$ $\Omega$m      &    f$_\tn{th}$           &  10                  & \Gy I$_\tn{c}$      & \Gy 50 $\mu$A      \\                                                
    J$_\tn{sot}$      &  1.5$\times$10$^\tn{10}$ Am$^\tn{-2}$   &    sf$_\tn{area}$        &  4                   & \Gy I$_\tn{sot}$    & \Gy 50 $\mu$A      \\                                                   
    $\rho_\tn{sot}$   &  6.7$\times$10$^\tn{-6}$ $\Omega$m      &    fanout                &  10                  & \Gy R$_\tn{sc}$     & \Gy 600 $\Omega$   \\                                                
    l$_\tn{min}$      &  30 nm                                  &    \Gy k$_\tn{SuperMag}$ &  \Gy 1.12            & \Gy R$_\tn{sot}$    & \Gy 60 $\Omega$    \\                     
    w$_\tn{min}$      &  30 nm                                  &    \Gy w                 &  \Gy 335 nm          &                     &                    \\                     
    th$_\tn{sot}$     &  10 nm                                  &    \Gy l                 &  \Gy 30 nm           &                     &                    \\                     
     \hline                                                      
     
\end{tabular}
\label{tab:sm_params}   

\end{table}

\textit{Area -- } Area for a SuperMag switch is computed as $wl$.  Additionally, an overhead of $3wl$ per switch is anticipated to be required to allow for routing.  Area per cell is therefore $A_{cell} = 4wlN$ where $N$ is the number of switches in the cell.  As with the CMOS and RSFQ systems, standard cell counts are reported in the genus cell report enabling area estimations.

\textit{Power -- } Genus is configured to report off-state superconductor leakage in terms of the power dissipated by the sensing wire of the OFF-SuperMag switch in a single inverter.  The total P$_\tn{sc,total}$ is therefore the leakage reported by genus multiplied by the unit power, P$_\tn{sc,inv}$, computed as described Supplementary Note \ref{supnote:sm_calcs}.  Genus also reports the total switching energy per clock cycle as ``internal power'' in terms of the energy dissipated per switching event, E$_\tn{sw,unit}$, which is calculated as described in Supplementary Note \ref{supnote:sm_calcs}.  Total switching power is therefore computed as the genus-computed average number of switching events per clock cycle multiplied by E$_\tn{sw,unit}$ and the clock frequency.  P$_\tn{sot,total}$ is extracted from the cell report.  Assuming no gate-level clocking scheme is used, SOT current is driven continuously through \textit{every} SuperMag switch of every combinational gate and \textit{four} SuperMag switches of every DFF.  The total number of switches contributing to P$_\tn{sot,total}$ is therefore N$_\tn{comb}$ $+$ 4 $\times$ N$_\tn{DFF}$ where N$_\tn{comb}$ is the number of combinational switches, and N$_\tn{DFF}$ is the number of DFF standard cells.  P$_\tn{sot,total}$ is therefore computed by multiplying this number by P$_\tn{sot,fo10}/$10, with P$_\tn{sot,fo10}$ computed as described in the previous Supplementary Note.

\textit{Delay -- } Delay is computed as the minimum clock period of the design, which is computed as the maximum number of combinational stages in a timing path multiplied by the SuperMag switching time.  As SuperMag signals are current based, no delay is caused by capacitive charging.

\textit{RAM -- } RAM decoder area and energy are described as above. RAM array area is computed as the bit cell area (four SuperMag switches) times 16,384 (2 KB).  RAM array static power is zero as RAM is non-volatile.

\end{document}